\documentstyle[12pt]{article}

\skewchar\fivmi='177
\skewchar\sixmi='177
\skewchar\sevmi='177
\skewchar\egtmi='177
\skewchar\ninmi='177
\skewchar\tenmi='177
\skewchar\elvmi='177
\skewchar\twlmi='177
\skewchar\frtnmi='177
\skewchar\svtnmi='177
\skewchar\twtymi='177
\def\@magscale#1{ scaled \magstep #1}
\skewchar\fivsy='60
\skewchar\sixsy='60
\skewchar\sevsy='60
\skewchar\egtsy='60
\skewchar\ninsy='60
\skewchar\tensy='60
\skewchar\elvsy='60
\skewchar\twlsy='60
\skewchar\frtnsy='60
\skewchar\svtnsy='60
\skewchar\twtysy='60

\catcode`@=11
\def\un#1{\relax\ifmmode\@@underline#1\else
        $\@@underline{\hbox{#1}}$\relax\fi}
\catcode`@=12

\def\a{\alpha}
\def\b{\beta}

\def\d{\delta}
\def\e{\epsilon}

\def\g{\gamma}

\def\k{\kappa}
\def\l{\lambda}
\def\m{\mu}
\def\n{\nu}
\def\o{\omega}
\def\p{\pi}
\def\q{\theta}
\def\r{\rho}
\def\s{\sigma}
\def\t{\tau}

\def\x{\xi}
\def\z{\zeta}

\def\F{\Phi}
\def\G{\Gamma}

\def\L{\Lambda}

\def\Q{\Theta}
\def\S{\Sigma}

\def\X{\Xi}

\def\ch{{\cal H}}

\def\dslash{\not{\hbox{\kern-2pt $\partial$}}}
\def\Dslash{\not{\hbox{\kern-4pt $D$}}}
\def\pslash{\not{\hbox{\kern-2.3pt $p$}}}
 \newtoks\slashfraction
 \slashfraction={.13}
 \def\slash#1{\setbox0\hbox{$ #1 $}
 \setbox0\hbox to \the\slashfraction\wd0{\hss \box0}/\box0 }

\font\ro=cmsy10                          
\def\kcr{{\hbox{\ro \char'170}}}                
\def\ktl{{\hbox{\ro \char'170}}}        
\def\ktr{{\hbox{\ro \char'170}}}        
\def\kbl{{\hbox{\ro \char'170}}}        
\def\kbr{{\hbox{\ro \char'170}}}        

\def\bo{{\raise.15ex\hbox{\large$\Box$}}}               
\def\pr{\prod}                                          
\def\TH{{\raise.2ex\hbox{$\displaystyle \bigodot$}\mskip-4.7mu \llap H \;}}
\def\face{{\raise.2ex\hbox{$\displaystyle \bigodot$}\mskip-2.2mu \llap {$\ddot
        \smile$}}}                                      

\def\sp#1{{}^{#1}}                              
\def\leftrightarrowfill{$\mathsurround=0pt \mathord\leftarrow \mkern-6mu
        \cleaders\hbox{$\mkern-2mu \mathord- \mkern-2mu$}\hfill
        \mkern-6mu \mathord\rightarrow$}
\def\dvec#1{\vbox{\ialign{##\crcr
        \leftrightarrowfill\crcr\noalign{\kern-1pt\nointerlineskip}
        $\hfil\displaystyle{#1}\hfil$\crcr}}}           

\def\frac#1#2{{\textstyle{#1\over\vphantom2\smash{\raise.20ex
        \hbox{$\scriptstyle{#2}$}}}}}                   
\def\sfrac#1#2{{\vphantom1\smash{\lower.5ex\hbox{\small$#1$}}\over
        \vphantom1\smash{\raise.4ex\hbox{\small$#2$}}}} 
\def\bfrac#1#2{{\vphantom1\smash{\lower.5ex\hbox{$#1$}}\over
        \vphantom1\smash{\raise.3ex\hbox{$#2$}}}}       
\def\afrac#1#2{{\vphantom1\smash{\lower.5ex\hbox{$#1$}}\over#2}}    

\newskip\humongous \humongous=0pt plus 1000pt minus 1000pt
\def\caja{\mathsurround=0pt}

\newif\ifdtup
\def\panorama{\global\dtuptrue \openup2\jot \caja
        \everycr{\noalign{\ifdtup \global\dtupfalse
        \vskip-\lineskiplimit \vskip\normallineskiplimit
        \else \penalty\interdisplaylinepenalty \fi}}}
\def\eqalignno#1{\panorama \tabskip=\humongous                  
        \halign to\displaywidth{\hfil$\displaystyle{##}$
        \tabskip=0pt&$\displaystyle{{}##}$\hfil
        \tabskip=\humongous&\llap{$##$}\tabskip=0pt
        \crcr#1\crcr}}

\def\ref#1{$\sp{#1)}$}

\def\np#1{{\it Nucl. Phys.} {\bf B#1}}

\def\prl#1{{\it Phys. Rev. Letts.} {\bf #1}}
\def\pr#1{{\it Phys. Rev.} {\bf #1}}
\def\pl#1{{\it Phys. Letts.} {\bf #1B}}

\parskip=\medskipamount                 
\thicklines                         

\def\oldheadpic{                                
        \setlength{\unitlength}{.4mm}
        \thinlines
        \par
        \begin{picture}(349,16)
        \put(325,16){\line(1,0){4}}
        \put(330,16){\line(1,0){4}}
        \put(340,16){\line(1,0){4}}
        \put(335,0){\line(1,0){4}}
        \put(340,0){\line(1,0){4}}
        \put(345,0){\line(1,0){4}}
        \put(329,0){\line(0,1){16}}
        \put(330,0){\line(0,1){16}}
        \put(339,0){\line(0,1){16}}
        \put(340,0){\line(0,1){16}}
        \put(344,0){\line(0,1){16}}
        \put(345,0){\line(0,1){16}}
        \put(329,16){\oval(8,32)[bl]}
        \put(330,16){\oval(8,32)[br]}
        \put(339,0){\oval(8,32)[tl]}
        \put(345,0){\oval(8,32)[tr]}
        \end{picture}
        \par
        \thicklines
        \vskip.2in}
\def\oldtitle#1#2#3#4{\oldheadpic\begin{center}\vglue.5in{\large\bf #1}\\[.6in]
        {#2}\\[.1in] {\it Department of Physics and Astronomy}\\
        {\it University of Maryland, College Park, MD 20742}\\[.6in]
        Physics Publication \#{#3}\\ {#4}\\[1.5in] {\bf ABSTRACT}\\[.1in]
        \end{center} \begin{quotation}}                 
\def\oldTitle#1#2#3#4#5#6#7{\oldheadpic\begin{center} \vglue .4in
        {\large\bf #1}\\[.4in]
        {#2}\\[.1in] {\it Department of Physics and Astronomy}\\
        {\it University of Maryland, College Park, MD 20742}\\[.1in]
        {#3}\\[.1in] {\it {#4}}\\ {\it {#5}}\\[.4in]
        Physics Publication \#{#6}\\ {#7}\\[.5in] {\bf ABSTRACT}\\[.1in]
        \end{center} \begin{quotation}}                 
\def\border{                                            
        \setlength{\unitlength}{1mm}
        \newcount\xco
        \newcount\yco
        \xco=-21
        \yco=12
        \begin{picture}(140,0)
        \put(\xco,\yco){$\ktl$}
        \advance\yco by-1
        {\loop
        \put(\xco,\yco){$\kcr$}
        \advance\yco by-2
        \ifnum\yco>-240
        \repeat
        \put(\xco,\yco){$\kbl$}}
        \xco=158
        \yco=12
        \put(\xco,\yco){$\ktr$}
        \advance\yco by-1
        {\loop
        \put(\xco,\yco){$\kcr$}
        \advance\yco by-2
        \ifnum\yco>-240
        \repeat
        \put(\xco,\yco){$\kbr$}}
        \put(-20,13){\tiny University of Maryland Elementary Particle
Physics University of Maryland Elementary Particle Physics University of
Maryland Elementary Particle Physics}
        \put(-20,-241.5){\tiny University of Maryland Elementary
Particle Physics University of Maryland Elementary Particle Physics
University of Maryland Elementary Particle Physics}
        \end{picture}
        \par\vskip-8mm}
\def\bordero{                                           
        \setlength{\unitlength}{1mm}
        \newcount\xco
        \newcount\yco
        \xco=-31
        \yco=12
        \begin{picture}(140,0)
        \put(\xco,\yco){$\ktl$}
        \advance\yco by-1
        {\loop
        \put(\xco,\yco){$\kclr}
        \advance\yco by-2
        \ifnum\yco>-240
        \repeat
        \put(\xco,\yco){$\kbl$}}
        \xco=151
        \yco=12
        \put(\xco,\yco){$\ktr$}
        \advance\yco by-1
        {\loop
        \put(\xco,\yco){$\kcr$}
        \advance\yco by-2
        \ifnum\yco>-240
        \repeat
        \put(\xco,\yco){$\kbr$}}
        \put(-20,12){\ooo
bacdefghidfghghdhededbihdgdfdfhhdheidhdhebaaahjhhdahba

hgdedge
   hgfdiehhgdigicba}
        \put(-20,-241.5){\ooo
ababaighefdbfghgeahgdfgafagihdidihiidhiagfedhadbfd

ecdcdfa
   gdcbhaddhbgfchbgfdacfediacbabab}
        \end{picture}
        \par\vskip-8mm}
\def\headpic{                                           
        \indent
        \setlength{\unitlength}{.4mm}
        \thinlines
        \par
        \begin{picture}(29,16)
        \put(165,16){\line(1,0){4}}
        \put(170,16){\line(1,0){4}}
        \put(180,16){\line(1,0){4}}
        \put(175,0){\line(1,0){4}}
        \put(180,0){\line(1,0){4}}
        \put(185,0){\line(1,0){4}}
        \put(169,0){\line(0,1){16}}
        \put(170,0){\line(0,1){16}}
        \put(179,0){\line(0,1){16}}
        \put(180,0){\line(0,1){16}}
        \put(184,0){\line(0,1){16}}
        \put(185,0){\line(0,1){16}}
        \put(169,16){\oval(8,32)[bl]}
        \put(170,16){\oval(8,32)[br]}
        \put(179,0){\oval(8,32)[tl]}
        \put(185,0){\oval(8,32)[tr]}
        \end{picture}
        \par\vskip-6.5mm
        \thicklines}
\def\title#1#2#3#4{\border\headpic {\hbox to\hsize{#4 \hfill UMDEPP #3}}\par
        \begin{center} \vglue .5in {\large\bf #1}\\[.6in]
        {#2}\\[.1in] {\it Department of Physics and Astronomy}\\
        {\it University of Maryland, College Park, MD 20742}\\[1.5in]
        {\bf ABSTRACT}\\[.1in] \end{center} \begin{quotation}}  
\def\Title#1#2#3#4#5#6#7{\border\headpic
        {\hbox to\hsize{#7 \hfill UMDEPP #6}}\par
        \begin{center} \vglue .4in {\large\bf #1}\\[.4in]
        {#2}\\[.1in] {\it Department of Physics and Astronomy}\\
        {\it University of Maryland, College Park, MD 20742}\\[.1in]
        {#3}\\[.1in] {\it {#4}}\\ {\it {#5}}\\[.5in] {\bf ABSTRACT}\\[.1in]
        \end{center} \begin{quotation}}                 
\def\endtitle{\end{quotation}\newpage}                  

\def\sect#1{\bigskip\medskip \goodbreak \noindent{\bf {#1}} \nobreak \medskip}
\def\refs{\sect{REFERENCES} \footnotesize \frenchspacing \parskip=0pt}
\def\Item{\par\hang\textindent}

\begin{document}
\def\sech{{\rm sech}~\!{r}}
\def\csch{{\rm csch}~\!{r}}
\def\css{{\rm csch}^2r}
\def\sss{{\rm sech}^2r}
\def\ssss{{\rm sech}^4r}
\def\th{{\rm tanh}~\!{r}}
\def\ths{{\rm tanh}^2r}
\def\chs{{\rm coth}^2r}
\def\qsech{{\rm sech}^3r}     %
\def\ch{{\rm coth}~r}
\def\qch{{\rm coth}^3r}       %
\def\x{\delta\!{f_0}}
\def\y{\delta\!{f_1}}
\def\z{\delta\Phi}
\def\gfrac#1#2{\frac {\scriptstyle{#1}}
        {\mbox{\raisebox{-.6ex}{$\scriptstyle{#2}$}}}}
\def\gg{{\hbox{\sc g}}}
\border\headpic {\hbox to\hsize{
October 1992 \hfill {UMDEPP 93-061}}}
{\hbox to\hsize{\hfill {FERMI-PUB-92/272-T}}}\par
\setlength{\oddsidemargin}{0.3in}
\setlength{\evensidemargin}{-0.3in}
\begin{center}
\vglue .08in
{\large\bf All or Nothing:\\ On the Small Fluctuations of Two--Dimensional
String--Theoretic Black Holes}
\\[.16in]
Gerald Gilbert\footnote {Research supported in part by NSF grant
\# PHY--91-19745}
\\[.02in]
{\it Center for Theoretical Physics and Department of Physics\\
University of Maryland at College Park\\
College Park, MD 20742 USA}\\
{\bf {\tt gng@umdhep.umd.edu}}\\[.04in] and\\[.04in]
Eric Raiten\\[.02in]
{\it Theory Group, MS 106\\
Fermi National Accelerator Laboratory\footnote {Operated by Universities
Research Association Inc. under contract with the U.S. D.O.E.}\\
P.O. Box 500, Batavia, IL 60510 USA}\\ {\bf {\tt
raiten@fnal.fnal.gov}}\\[.1in]

{\bf ABSTRACT}\\[.002in]
\end{center}
\begin{quotation}
{A comprehensive analysis of small fluctuations about two--dimensional
string--theoretic and string--inspired black holes is presented. It is shown
with specific examples that two--dimensional black holes behave in a radically
different way from all known black holes in four dimensions. For both the
$SL(2,R)/U(1)$ black hole and the two--dimensional black hole coupled to a
massive dilaton with constant field strength, it is shown that there are a
{\it continuous infinity} of solutions to the linearized equations of motion,
which are such that it is impossible to ascertain the
classical linear response. It is further shown that the
two--dimensional black hole coupled to a massive, linear dilaton
admits {\it no small fluctuations at all}.
We discuss possible implications of our results for the
Callan--Giddings--Harvey--Strominger black hole.}

\endtitle

\sect{{\bf I. Overview}}

\sect{{\bf I.1 Introduction}}

\noindent{Seemingly insuperable problems
may sometimes be forced to yield their secrets if one considers instead simpler
problems which are appropriately chosen. Whether
or not such a strategy will succeed in a specific case
depends in the first instance on the proviso that the chosen, simpler problem
be sufficiently closely related to the problem of authentic interest.
The discovery by Stephen Hawking that, when considered in the semi--classical
approximation, black holes may emit thermally--distributed radiation has led
to a number of problems which have so far proved too difficult to solve.
The supreme example is the problem posed by the possibility that black holes,
in the event that they ``evaporate" absolutely and completely, may
irrevocably obliterate information {\it in principle}. If this is how
Nature behaves then Quantum Mechanics would seem to be unable to provide an
adequate account, for it is a {\it sine qua non} that the wave
function must display a well--defined unitary time development [1]. It is
clearly
premature to abandon quantum mechanics before giving its principles a fair
chance, and it is unlikely that this will have been achieved by relying solely
on the semi--classical approximation. The
discovery of black hole radiation was originally presented with the
understanding that one was considering solutions to Einstein's equations in
four dimensions, and it is in this context that it has so far proved too
difficult to go beyond the semi--classical approximation.}

\noindent{Recently a great deal of attention has been focussed on interesting
related developments in string theory. Black hole solutions have been
discovered to the still--unwritten equations of motion in string theory. One
may say that these solutions are ``unsatisfactory" in differing amounts and in
various ways: They are available only at the level of the Born approximation
in string theory; some of them are further approximate in that higher--order
contributions to the sigma model on the sphere have been neglected in their
derivation; some of them are actually solutions to a ``string--inspired"
theory and not to string theory; some of them are defined in a mythical
two--dimensional universe. Nevertheless, and with
specific regard to the last point, it is precisely because of the relative
simplicity which may be found in two--dimensional solutions that it is hoped
that their study may assist in the resolution of outstanding problems of
four--dimensional black holes.}

\noindent{The deep puzzles of black hole physics which one would like to
solve are in large part inherently quantum mechanical.
In the case of four--dimensional black holes one is at present able to say
much more about the classical mechanical behavior than about the quantum
mechanical behavior. It is hoped in particular that one may study the quantum
mechanical aspects of two--dimensional black holes (for which purpose their
origins in string theory are probably immaterial) and draw inferences
therefrom which may be successfully applied to the quantum mechanics of
four--dimensional black holes. As always, the correspondence principle must
serve as our guide in moving between quantum mechanics and classical
mechanics. In considering the possibly simpler quantum mechanics of
two--dimensional black holes to aid in the understanding of four--dimensional
black holes, it is natural to require that the correspondence limit of the
substitute, 2$d$ configuration behave in a reasonably similar way
to that of the 4$d$ configuration of authentic interest. In this paper
we perform the first comprehensive analysis of the small fluctuations of
specific two--dimensional string--theoretic and string--inspired black holes
which have been the focus of recent research.{\footnote {See Note 1 below
which follows the conclusion section of this article.}} We demonstrate that
these {\it two--dimensional black holes display classical behavior
which differs radically from that of all known four--dimensional black holes.}
We find this to be the case for the $SL(2,R)/U(1)$ black hole [2]
as well as for
two--dimensional black holes coupled to a massive dilaton. In the
cases of the Witten black hole and the black hole coupled to a massive dilaton
with constant field strength we find that the linearized equations of motion
admit a continuous infinity of solutions which are such that it is {\it in
principle} impossible to ascertain the classical linear response,
while we find that the black hole coupled to a massive linear dilaton admits
{\it no small fluctuations at all.} We may say therefore that the physics of
these two--dimensional black holes is an ``all or nothing" proposition.}

\noindent{It is an element of geometry that the Einstein--Hilbert lagrange
density in a two--dimensional theory of gravitation is a total divergence. It
is furthermore the case
that two--dimensional dilaton gravity is characterized by the
absence of propagating degrees of freedom. We note that it is
highly unlikely that the
unusual behavior displayed by the various two--dimensional black holes
studied in this paper is a
consequence of this fact. We see very different types of linear response
behavior for the various black hole examples we study, although they share
the absence of
propagating degrees of freedom. While the extremely unusual classical
behavior found
for specific two--dimensional black holes is not fully understood as
to its origin, we may speculate on the possible implications of these results
for other two--dimensional black holes. In particular, we discuss the
so--called CGHS black hole [3],
which is closely related to the Witten black hole,
and which is being studied in an attempt to secure a better understanding
of the physics of four--dimensional black holes. Based on the results of our
analysis of different types of two--dimensional black holes, it
would not be surprising to discover that the CGHS black hole too
displays radically different classical linear response behavior from the known
black holes in four dimensions. It is essential to repeat the calculation of
the present paper for this configuration in order to determine if it
is reasonable to expect that correct inferences applicable to
four--dimensional black holes can be drawn from its study.}

\noindent{This report is organized as follows. The remainder of the
Section I is devoted to a review of the recent work that has been done on
two--dimensional black holes in Section I.2, after which we provide
a pr\'ecis of the group--theoretic derivation of the $SL(2,R)/U(1)$ black
hole in Section I.3. In Section II we
perform the analysis of the small fluctuations of
black holes: in Section II.1 we provide a general description of the technique
for black holes in arbitrary dimensions which will be useful to those who are
not familiar with this subject; in Section II.2 we specialize the analysis to
the
case of two dimensions by first providing an account of the general formulae
relevant to two--dimensional theories of gravitation in Section II.2.a, after
which we consider comprehensively in turn the $SL(2,R)/U(1)$
black hole in Section II.2.b, the black hole coupled to a massive dilaton with
constant field strength in Section II.2.c.i and the black hole coupled to a
massive
linear dilaton in Section II.2.c.ii. We present our conclusions in Section
III, which
is followed by a section of Notes detailing certain technical points, Tables
of numerical results and an Appendix.}

\sect{{\bf I.2 Review of Related Work}}

\noindent{In this section we present a brief survey of the
recent research efforts devoted to two--dimensional black holes, and in
the next section we review the group--theoretic derivation of the
$SL(2,R)/U(1)$ black hole, both of which
will be useful to those who are not familiar with this subject. Experts may
proceed directly to the analysis of the linearized equations of motion in
Section II.
For the particular case of two--dimensional black holes,
a great deal of research has followed the observation by Witten [2] that the
conformal field theory based on the non--compact coset model $SL(2,R)/U(1)$,
which had been developed by Bars and Nemeschansky [4],
Rocek,{\it et. al.} [5], and others, consists of
a two--dimensional black hole coupled to the dilaton. More importantly, the
asymptotic form of the metric
is just the linear dilaton vacuum which is studied in the $c=1$
matrix model.  Furthermore, the endpoint of the Hawking radiation process,
{\it i.e.}, the $M\rightarrow 0$ limit, where $M$ is the mass of the black
hole, also approaches the linear dilaton vacuum.}

\noindent{Using the algebraic structures inherent in the $G/H$ construction
of this model, a number of groups,
including Dijkgraaf {\it et. al.} [6], Distler
and Nelson [7], and Chaudhuri and Lykken [8],
have considered the spectrum of states
and their correlation functions.   In particular, Chaudhuri and Lykken [8]
emphasize the $W_{\infty}$--like structure of the model's marginal operators,
to which point we shall return in a later section.}

\noindent{Among other developments, Bars [9], and Ginsparg and
Quevedo [10]
have classified all $G/H$ models which give rise to spacetimes with only a
single time--like coordinate, in any number of dimensions.
In addition to the obvious physical
importance of having only a single time--like coordinate, it is argued
by Bars [9] that models with more than one time-like coordinate are
likely to be ill--behaved, since the Virasoro conditions (or
equivalently, light--cone gauge) are generally sufficient to remove
the negative norm states generated by only a single time--like coordinate.
A Hamiltonian formalism is developed, in which the
target space metric, antisymmetric tensor and dilaton are determined
to {\it all} orders in $\a '$.
Ginsparg and Quevedo [10] have stressed the connection between target space
singularities and fixed points of the gauge transformation generated
by $H$.  Gibbons and Perry [11] have discussed the thermodynamics of the
$SL(2,R)/U(1)$ solution and related heterotic solutions.}

\noindent{Another model which has recently attracted great interest is the
dilaton gravity model of Callan, Giddings, Harvey and Strominger (CGHS) [3].
The study of the model begins with the so--called ``string--inspired" action,
to which a set of minimally--coupled free scalar fields is added. In the
initial model, they found
that any scalar wave impinging on the linear dilaton vacuum creates a
black hole. Calculating the Hawking radiation  via its relation in
two dimensions to the trace anomaly, one finds a divergent integrated flux.
The resolution to this apparent dilemma lies in the neglect of backreaction
on the metric. Therefore, CGHS modified their action to include the one--loop
effects of the scalar fields.}

\noindent{While the initial hopes that the Hawking
radiation could then be treated well within the
semi--classical regime were
later proven false [12], a number of groups continue to investigate the
detailed behavior of the model. DeAlwis [13],
as well as Bilal and Callan [14], have attempted
to quantize the system by a Distler--David--Kawai
approach (see also Hamada [15]). That is to say,
they try to form a non--linear sigma model which solves
the appropriate beta--function
equations and reduces to the CGHS model in the semi--classical limit.
As pointed out by Giddings and Strominger [16], such models generally do not
have a well--defined ground state.  They point out an ambiguity in the
regularization of the path integral of the theory, with the result that
essentially an infinite number of counterterms must be specified even though
dilaton gravity is renormalizable.
In other work, Hawking and Stewart [17] claim
numerical evidence that the CGHS black hole will end in a ``thunderbolt",
{\it i.e.}, a singularity which propagates out to infinity on a spacelike or
null path.  Russo, Susskind and Thorlacius [18] discuss models
in which a naked singularity forms,
but claim that appropriate boundary conditions can be imposed which will
prevent the loss of any quantum mechanical information.}

\noindent{Variations of these models have been treated recently by a number
of authors. One variation with which we will be concerned here are models
with a nonvanishing dilaton potential, considered recently by Gregory
and Harvey [19],
and by Horne and Horowitz  [20] (the latter in four dimensions only).
Others include charged and supersymmetric black holes [21].}

\noindent{In spite of all these efforts, many of the central questions
concerning
both black hole physics and nonperturbative string backgrounds remain
essentially unanswered.  The proper quantization of the CGHS model is needed
in order to probe the problems of information loss and the endpoint of
Hawking radiation, but even the full set of classical solutions of the
model are not known.  Starting from the string--theoretic $SL(2,R)/U(1)$
model, one faces a similar problem, in that generally one is
only able to perform calculations in the semi-classical limit.}

\sect{{\bf I.3 Review of Group--Theoretic Derivation of the $SL(2,R)/U(1)$
Black Hole}}

\noindent{The first black hole we will consider is the $SL(2,R)/U(1)$ model,
discovered in various forms by Witten [2], Mandal {\it et. al.} [22], and Bars
{\it et. al.} [4]. Here we briefly review its construction, generally
following the notation of Witten [2].}

\noindent{We begin with the ungauged $SL(2,R)$ Wess--Zumino--Witten (WZW)
action}
$$S_{WZW}={k\over {8\p}}\int_{\S}d^2z~\sqrt{h}h^{ij}
{\rm Tr}(g^{-1}\partial_ig\ g^{-1}
\partial_jg)+ik\G ,\eqno(1)$$
where $\S$ is a Riemann surface with metric $h$, $g$ is an $SL(2,R)$-valued
field on $\S$, and $k$ is real and positive.  $\G$ is the Wess--Zumino
term, which is usually represented as
$$\G ={1\over {12\p}}\int_Bd^3y~{\rm Tr}
(g^{-1}
dg\wedge g^{-1}dg\wedge g^{-1}dg),\eqno(2)$$
where $B$ is a three--dimensional manifold with boundary equal to $\S$.
In this expression, $g$ has been extended from a field on $\S$ to a field
on $B$, but $\G$ is independent of this choice.

\noindent{The Euclidean version of the black hole is now obtained by
gauging the $U(1)$ subgroup the infinitesimal action of which is given by}
$$\d g=\e\{{\cal G}g+g{\cal G}\},\eqno(3)$$
where ${\cal G}$ is the constant $SL(2,R)$ element
$${\cal G}=\pmatrix{0&1\cr -1&0\cr}.\eqno(4)$$
To gauge this symmetry, we introduce a gauge field $A$ with the
transformation law
$$\d A_i=-\partial_i\e .\eqno(5)$$
In local complex coordinates $z,\bar z$, the
gauge invariant action now takes the form
$$S=S_{WZW}+{k\over {2\p}}\int_{\S}d^2z
\{ A_{\bar z}{\rm Tr}({\cal G}g^{-1}\partial_z g)+
A_z{\rm Tr}({\cal G}\partial_{\bar z}g g^{-1})+
A_zA_{\bar z} (-2+{\rm Tr}({\cal G}g{\cal G}g^{-1}))\}.
\eqno(6)$$
One now fixes the gauge by setting
$$g={\rm cosh}~r+{\rm sinh}~r\pmatrix{{\rm cos}~\q&{\rm sin}~\q\cr
{\rm sin}~\q&-{\rm cos}~\q\cr}.\eqno(7)$$
The gauge field $A$ appears quadratically and without derivatives. Integrating
it out and dropping the Wess--Zumino term (as it is a total derivative)
one finds the effective action
$$I_0={k\over {4\p}}\int d^2x\sqrt{h}h^{ij}(\partial_ir\partial_jr+
\ths \partial_i\q\partial_j\q ).\eqno(8)$$
This has the form of a nonlinear sigma model with target space metric
$$ds^2={k\over 2}((dr)^2+\ths (d\q )^2).\eqno(9)$$

\noindent{It is well--known [23] that upon integrating out the gauge field
one finds that the integration measure yields a finite correction to
the action:}
$$I=I_0-{1\over {8\p}}\int d^2x\sqrt{h}\F (r,\t )R,\eqno(10)$$
where $R$ is the world sheet curvature and $\F$ is the target
space dilaton.  In the present case, one finds
$$\F =2\ {\rm ln\ cosh}~r+\eta,\eqno(11)$$
where $\eta$ is a constant related to the black hole mass.
This form of the
dilaton can also be seen from the target space action which we will consider
in the next section.

\noindent{The Lorentzian signature form of the black hole, which we shall
use in the next section, can be obtained most simply by the analytic
continuation $\q\rightarrow it$, or by gauging a different $U(1)$ subgroup,
in which the matrix ${\cal G}$ above is replaced by}
$${\cal G}\rightarrow \pmatrix{1&0\cr0&-1\cr}.\eqno(12)$$

\noindent{As noted by Witten [2], if one computes the
central charge from this action, it
differs from the $SL(2,R)/U(1)$ value of $2+{6\over {k-2}}$ by an amount
of order
${1\over {k^2}}$, implying that there are further corrections from the
integration over
the gauge field.
These corrections would presumably appear in terms higher order in the
sigma model coupling $\alpha '$.  We will return to this point in a later
section.}

\noindent{Nearly all of the 2--d black holes in the recent literature are
related in some way to the $SL(2,R)/U(1)$ black hole. For example, the
analysis of the CGHS model [3] begins
with the $M=0$ limit of the $SL(2,R)/U(1)$
black hole
(which corresponds to $\eta\rightarrow -\infty$).
A set of minimally--coupled scalars is added to the action, and one finds
that {\it any} incoming scalar wave creates a black hole.  Of course, for a
given incoming scalar distribution, it is not known whether the resulting
background solution corresponds to a conformal field theory.}

\noindent{The
massive dilaton models recently considered by Gregory and Harvey [19] are also
related to the $SL(2,R)/U(1)$ model, in that
they are solutions of the same target space action, but with the addition
of an explicit potential for the dilaton (though they do not contain the
scalars of the CGHS model). By taking the mass to zero, one can recover
the $SL(2,R)/U(1)$ model.  Of course, the mass terms imply that
these models definitely do not correspond to a conformal field theory.
While such terms do
not appear in string perturbation theory, it is widely speculated that they are
related to supersymmetry breaking.  Furthermore,
a mass must be generated since the dilaton is related
to the string coupling constant.  Experimental tests of conventional
Brans--Dicke models also put tight constraints on very light scalars,
though there are recent models in which such limits are evaded if the metric
is chosen to couple differently to a
``dark matter'' dilaton than to ordinary
``visible'' matter [24].}

\sect{{\bf II. Analysis of Linearized Equations}}
\sect{{\bf II.1 Small--Fluctuation Analyses of Black Holes}}

\noindent{In the following sections we shall explicitly analyze the
perturbations
of two--dimensional string--theoretic black holes. Here we shall first survey
the general procedure used in the analysis of the perturbations of black holes
in any number of dimensions [25]. We suppose that one has found a black hole
solution to the coupled field equations of an interacting
system consisting of gravitation and, in general, additional ``matter" fields
of various possible types, including different spins. The
different species of ``matter" will be denoted by labels $\psi^{(1)}$ through
$\psi^{(n)}$, where possible tensor indices have been suppressed. The field
configuration which defines the black hole solution, which we will refer to as
the {\it background}, will be denoted by the collection of
$g_{\m\n}^B$ and $\psi_B^{(i)}$.
The coupled field equations to which the background provides a solution are
then given by}

$$R_{\m\n}=T_{\m\n}^{(1)}+\cdots+T_{\m\n}^{(n)} ~,\eqno(13)$$
$$\eqalignno{{\hat {\cal H}}^{(1)}\left(\psi^{(1)}\right)&={\cal I}^{(1)},\cr
&\vdots\cr {\hat {\cal H}}^{(n)}\left(\psi^{(n)}\right)&={\cal I}^{(n)}
{}~,&(14)\cr}$$

\noindent{where the $T_{\m\n}^{(i)}$ are the various stress tensors
associated with the different ``matter" fields, the ${\hat {\cal H}}^{(i)}$
are in general coupled, nonlinear, tensor--valued, second--order partial
differential operators which may depend on the different fields and the
${\cal I}^{(i)}$ are possible source terms. From these coupled nonlinear
equations one now computes the associated
first--order variations, which yields\footnote{It is important to note
that there is a proper order in which to perform these computations: it is
only {\it after} calculating the abstract variations that one may substitute
the background field values into eqs. (15) through (16). If this order
is not respected one will in general miss those terms which vanish in the
background but do not fluctuate to zero.}}

$$\d R_{\m\n}=\d T_{\m\n}^{(1)}+\cdots+\d T_{\m\n}^{(n)} ~,\eqno(15)$$
$$\eqalignno{0=&\d\left[{\hat {\cal H}}^{(1)}\left(\psi^{(1)}\right)\right]
-\d{\cal I}^{(1)},\cr &\vdots\cr 0=&\d\left[{\hat
{\cal H}}^{(n)}\left(\psi^{(n)}\right)\right]
-\d{\cal I}^{(n)} ~.&(16)\cr}$$

\noindent{One next substitutes the background field values $g_{\m\n}^B$ and
$\psi_B^{(i)}$ into
these equations and then works out the reduction of the system which results
upon identifying any integrability conditions and imposing any kinematical
constraints. This leads to the following system of {\it linear}, coupled
partial--differential equations}

$$\eqalignno{{\hat \Q}^{(1)}\y&=\X^{(1)},\cr &\vdots\cr {\hat \Q}^{(m)}
\d f_m&=\X^{(m)} ~,&(17)\cr}$$

\noindent{where the $\d f_i$ are the distinct perturbations of the
background fields and the ${\hat \Q}^{(i)}$ are linear partial--differential
operators which depend on the background but are independent of the various
perturbations. In these equations, for a given $\d f_i$ the corresponding
$\X^{(i)}$ is a function of as many as $m-1$ of the {\it remaining}
perturbations and their derivatives and in general one has $m\le n$. In the
subsequent treatment of these equations one usually assumes that all field
perturbations have a time--dependence $\propto e^{i\o t}$, where $\o$ is a
non--dispersive frequency, and a temporal Fourier analysis is performed.}

\noindent{One next looks for an appropriate separation of variables in
order to transform eqs. (17) into a set of coupled ordinary differential
equations. The chosen separation must be consistent with the boundary
conditions imposed on the field perturbations.
It is then usually convenient to introduce integrating factors
which serve to eliminate all first derivative terms, after which one attempts
to decouple the resulting system of ordinary differential equations in two
steps. One first searches for a transformation of the dependent variables
which will allow the system to be expressed in the form:}

$$\pmatrix{{\hat D}^2&~&~\cr ~&\ddots&~\cr ~&
{}~&{\hat D}^2\cr}\pmatrix{\y\cr \vdots\cr \d f_m\cr}=\pmatrix{
P_{11}&\cdots&
P_{1m}\cr\vdots&\ddots&\vdots\cr P_{m1}&\cdots &P_{mm}\cr}\pmatrix{\y\cr
\vdots\cr \d f_m\cr}=\pmatrix{{\cal P}_1\cr\vdots\cr {\cal P}_m\cr}
{}~,\eqno(18)$$

\noindent{where ${\hat D}^2=d^2+\o^2$ (here $d$ is the spatial derivative),
the $P_{ij}$ are scalar functions and the ${\cal P}_i={\cal P}_i
\left(\y,\ldots,\d f_ m\right)$ are therefore in general linear functions of
all the distinct perturbations, but not of their derivatives. We say that the
system in this form has been only {\it differentially decoupled}. In the
second step we diagonalize the matrix $\left(P_{ij}\right)$, after which the
completely decoupled system of equations may be expressed in the form}

$${\hat D}^2\d p_i=v_i\d p_i ~,\eqno(19)$$

\noindent{where the physical perturbation functions $\d p_i$ are linear
combinations of the $\d f_i$ appropriate to the diagonalization of $\left(
P_{ij}\right)$, and the scalar functions $v_i$ are the
{\it perturbation potentials} which surround the black hole as a consequence
of the presence of the small fluctuations. Thus the original system has been
reduced to a set of completely decoupled Schr\"odinger--like radial equations.
As a result, once one has worked out the
explicit expressions for the $v_i$ it is possible to study the properties of
any possible bound states, to calculate the various scattering coefficients
associated with different incident perturbations and
in general to determine completely the linear response of the black hole to
diverse types of incoming waves of small to moderate intensity. It must be
emphasized, however, that there is no guarantee that it will be possible in
all cases to secure a suitable transformation of the dependent variables which
will allow the system of equations to be decoupled. Indeed, in the general
case this is an extremely challenging mathematical problem, and as we shall
see, it is in precisely this regard that two--dimensional black holes display
unexpected properties.}

\sect{{\bf II.2 Analysis of the Linearized Equations of Motion in Two
Dimensions}}
\sect{{\bf II.2.a General Formulae for Two--Dimensional Theories of
Gravitation}}

\noindent{In considering the small fluctuations of two--dimensional black
holes we first note that the most sufficiently general form for the
perturbed metric associated with a given initial configuration can be
represented by a diagonal matrix. This is always possible to arrange through a
transformation of the coordinates, as a result of which we note that we will
not encounter the analogues of the ``axial" perturbations which arise in the
study of black hole perturbations in more than two dimensions. The
first--order perturbations of two--dimensional black holes are entirely
``polar", and thus the metric tensor corresponding to the squared line
element}

$$ds^2=-e^{2f_0}dt^2+e^{2f_1}dr^2 ~,\eqno(20)$$

\noindent{will experience perturbations in the form}

$$\pmatrix{-e^{2f_0}&0\cr 0&e^{2f_1}\cr}~~\rightarrow ~~\pmatrix{
-e^{2f_0+2\d f_0}&0\cr 0&e^{2f_1+2\d f_1}\cr} ~.\eqno(21)$$

\noindent{We note in passing that, having taken the metric tensor to be
diagonal, and thus taken the perturbed metric tensor to be diagonal as well,
the choice of gauge in the perturbed system has been partially fixed. We shall
consider the residual gauge freedom in the perturbed system presently. We see
that with the reasonable assumption described in the previous
section that all field perturbations carry a time--dependence $\propto e^{i\o
t}$,\footnote{It is the case that the background spacetimes considered in
this paper are all characterized by a Killing vector. The norm of this vector
in the perturbed metric is {\it indefinite}, as a result of which
perturbations about the background are in general time--dependent.}
the various
equations for the different small fluctuations are {\it automatically}
separated in the coordinates. For any two--dimensional black hole, then, the
small fluctuations are determined by a system of coupled ordinary differential
equations.}

\noindent{We will consider the physics determined by the two--dimensional
action:{\footnote {We employ the sigma--model metric throughout the
following analysis.}}}

$$S=\left(2\pi\right)^{-1}\int d^2x{\sqrt{-g}}e^{-2\Phi}{\Big [}R+4\left(
\nabla\Phi\right)^2+4\L^2-e^{-2\Phi}V\left(\Phi\right){\Big ]} ~,
\eqno(22)$$

\noindent{where $\Phi$ is the dilaton field, $V$ is a generic ``potential"
for the dilaton, and $\L$ is the cosmological constant. Extremization of
the action with respect to the gravitational and dilaton fields, respectively,
leads to the following equations of motion:}

$$2e^{-2\Phi}{\Big \{}\nabla_\m\nabla_\n\Phi+g_{\m\n}{\Big [}\left(\nabla\Phi
\right)^2-\nabla^2\Phi-\L^2+{1\over 4}e^{-2\Phi}V\left(\Phi\right){\Big
]}{\Big \}}=0 ~,\eqno(23)$$
$$e^{-2\Phi}{\Big \{}R+4\L^2+4\nabla^2\Phi-4\left(\nabla\Phi\right)^2+
e^{-2\Phi}{\Big [}{1\over 2}{\partial V\over \partial\Phi}-2V\left(\Phi\right)
{\Big ]}{\Big \}}=0 ~.\eqno(24)$$

\noindent{Upon contracting both sides of eq.(23) with the metric tensor,
substituting the result into eq.(24), resolving the resulting equation into
components again and thereafter employing the convenient substitution
{\footnote {This substitution is in accord with the convention for the
relation between the string coupling and the exponential of the dilaton field
employed by Witten in [2]. It differs from the convention chosen
in [19,21],
which are devoted to the analysis of two-dimensional black holes coupled to
a massive dilaton. For these cases, when viewed as solutions to a
two--dimensional theory of gravitation, the choice is intrinsically
unimportant (in particular since there is no electromagnetic field present
and hence no duality transformation relating possible electric and magnetic
solutions),
and may be in any event irrelevant when viewed in the context of string theory
since the massive dilaton black holes considered in [19] may have little, and
possibly nothing whatsoever, to do with string theory.}} $\Phi\to -\Phi/2$
one may rewrite these equations as:}

$$0=\nabla^2\Phi+\left(\nabla\Phi\right)^2-4\L^2+e^\Phi\tilde V ~,\eqno(25)$$

\noindent{and}

$$R_{\m\n}=\nabla_\m\nabla_\n\Phi-{1\over 2}g_{\m\n}e^\Phi\left({1\over 2}
{\tilde V}'-\tilde V\right) ~,\eqno(26)$$

\noindent{where $\tilde V\equiv V\vert_{\Phi\to -\Phi/2}$ and a prime denotes
differentiation with respect to the dilaton field.}

\noindent{The variations of $\nabla^2\Phi$, $\left(\nabla\Phi\right)^2$ and
$\nabla_\m\nabla_\n\Phi$ are given by}

$$\eqalignno{\d\left(\nabla^2\Phi\right)=&\Phi_{,\n}\d g^{\m\n}_{~~,\m}+
{\Big [}\left(f_0+f_1\right)_{,\m}\Phi_{,\n}+\F_{\m\n}{\Big ]}\d g^{\m\n}+g^{
\m\n}\Phi_{,\n}\left(\x+\y\right)_{,\m}\cr &+g^{\m\n}\z_{,\m,\n}+{\Big [}
\left(f_0+f_1\right)_{,\m}g^{\m\n}+g^{\m\n}_{~~,\m}{\Big ]}\z_{,\n}
{}~,&(27)\cr}$$

$$\d\left[\left(\nabla\Phi\right)^2\right]=\Phi_{,\m}\Phi_{,\n}\d g^{
\m\n}+g^{\m\n}\Phi_{,\n}\z_{,\m}+g^{\m\n}\Phi_{,\m}\z_{,\n} ~,\eqno(28)$$

$$\d\left(\nabla_\m\nabla_\n\Phi\right)=\z_{,\m,\n}-\Phi_{,\l}\d \G_{\m\n}^
\l-\G_{\m\n}^\l\z_{,\l} ~,\eqno(29)$$

\noindent{and thus we derive from eqs. (25) and (26) the following
linearized perturbation equations:}

$$\eqalignno{0=&g^{\m\n}\z_{,\m,\n}+{\Big [}g^{\m\n}_{~~,\n}+g^{\m\n}\left(f_0
+f_1\right)_{,\n}+2g^{\m\n}\Phi_{,\n}{\Big ]}\z_{,\m}+e^\Phi\left(\d\tilde V+
\tilde V\z\right)\cr
&+\Phi_{,\n}\d g^{\m\n}_{~~,\m}+{\Big [}\left(f_0+f_1\right)_{,\m}\Phi_{,\n}
+2\Phi_{,\m}\Phi_{,\n}{\Big ]}\d g^{\m\n}+g^{\m\n}\Phi_{,\n}\left(
\x+\y\right)_{,\m} ~,&(30)\cr}$$

\noindent{and}

$$\eqalignno{\d R_{\m\n}=&\z_{,\m,\n}-\Phi_{,\l}\d \G_{\m\n}^\l-\G_{
\m\n}^\l\z_{,\l}\cr &-{1\over 2}e^\Phi{\Bigg [}\left({1\over 2}{\tilde V}'-
\tilde V\right)\d g_{\m\n}+g_{\m\n}\left({1\over 2}\d{\tilde V}'-\d\tilde V
\right)+g_{\m\n}\left({1\over 2}{\tilde V}'-\tilde V\right)\z{\Bigg ]}
{}~.\cr &&(31)\cr}$$

\noindent{We now note that for the general metric given by eq.(20)
computation reveals that the components of the Ricci tensor are given by:}

$$R_{00}=e^{2f_0-2f_1}\left(f_{0,r,r}+f_{0,r}^2-f_{0,r}f_{1,r}\right)
-\left(f_{1,0,0}+f_{1,0}^2-f_{0,0}f_{1,0}\right) ~,\eqno(32)$$

$$R_{01}=R_{10}=0 ~,\eqno(33)$$

\noindent{and}

$$R_{11}=-e^{-2f_0+2f_1}R_{00} ~.\eqno(34)$$

\noindent{One thus finds that the perturbations in the Ricci tensor are
given by:}

$$\eqalignno{\d R_{00}=&e^{2f_0-2f_1}{\Big [}\x_{,r,r}+\left(2f_{0,r}-
f_{1,r}\right)\x_{,r}-f_{0,r}\y_{,r}\cr &+2\left(\x-\y\right)\left(f_{
0,r,r}+f_{0,r}^2-f_{0,r}f_{1,r}\right){\Big ]}-\y_{,0,0}\cr &-\left(2f_{
1,0}-f_{0,0}\right)\y_{,0}+f_{1,0}\x_{,0} ~,&(35)\cr}$$

$$\d R_{01}=\d R_{10}=0 ~,\eqno(36)$$

\noindent{and}

$$\eqalignno{\d R_{11}=&-{\Big [}\x_{,r,r}+\left(2f_{0,r}-
f_{1,r}\right)\x_{,r}-f_{0,r}\y_{,r}{\Big ]}\cr &+e^{-2f_0+2f_1}{\Big [}
\y_{,0,0}+\left(2f_{1,0}-f_{0,0}\right)\y_{,0}-f_{1,0}\x_{,0}-2\left(
\x-\y\right)\left(f_{1,0,0}+f_{1,0}^2-f_{0,0}f_{1,0}\right){\Big ]}
{}~.\cr &&(37)\cr}$$

\sect{{\bf II.2.b The $SL(2,R)/U(1)$ Black Hole}}

\noindent{The $SL(2,R)/U(1)$ black hole with Lorentzian signature is the
solution to eqs. (25) and (26) for
$\tilde V={\tilde V}'=0$ given by the metric tensor of eq.(20) characterized
by the Wick rotation of the metric components given in eq.(9) above:}

$$g_{00}=-e^{2f_0}=-{k\over 2}\tanh^2r ~,\eqno(38)$$

$$g_{11}=e^{2f_1}={k\over 2} ~,\eqno(39)$$

\noindent{where $k=2\L^{-2}$ is the level of the underlying
Wess--Zumino action, along with a dilaton field given by}

$$\Phi=\log\cosh^2r+\eta ~,\eqno(40)$$

\noindent{where $\eta$ is a constant which is related to the mass $M$ of
the black hole through the equation $e^\eta=\left(k/2\right)^{1/2}M$.}

\noindent{In order to study the small fluctuations around this background
configuration we specialize eqs. (30) and (31) to the case of $\tilde V={
\tilde V}'=\d\tilde V=\d{\tilde V}'=0$, which yields:}

$$\eqalignno{0=&g^{\m\n}\z_{,\m,\n}+{\Big [}g^{\m\n}_{~~,\n}+g^{\m\n}\left(f_0
+f_1\right)_{,\n}+2g^{\m\n}\Phi_{,\n}{\Big ]}\z_{,\m}\cr
&+\Phi_{,\n}\d g^{\m\n}_{~~,\m}+{\Big [}\left(f_0+f_1\right)_{,\m}\Phi_{,\n}
+2\Phi_{,\m}\Phi_{,\n}{\Big ]}\d g^{\m\n}+g^{\m\n}\Phi_{,\n}\left(
\x+\y\right)_{,\m} ~,&(41)\cr}$$

\noindent{and}

$$\d R_{\m\n}=\z_{,\m,\n}-\Phi_{,\l}\d \G_{\m\n}^\l-\G_{\m\n}^\l\z_{,\l}
{}~.\eqno(42)$$

\noindent{The {\it basic perturbation equations} for the Witten black hole
are then obtained by substituting eqs. (38), (39) and (40) into eqs. (41)
and (42), which yields}

$$\x{''}+2\ch\ \x{'} -\sech\ \csch\ \y{'}
+\o^2\chs\y+\sech\ \csch\ \z{'}+\o^2\chs\ \z =0 \eqno(43)$$
$$\x{''}+2\sech\ \csch\ \x{'}-\th (\css +2)\y{'}+\o^2\chs\ \y+\z{''}=0
\eqno(44)$$
$$\z{''}+\th (\css +4)\z{'}+\o^2\chs\ \z-2\th\ \y{'}-8\y+2\th\ \x{'}=0
\eqno(45)$$
$$0=i\o\left(\z{'}-2\tanh r~\y-\sech~\csch~\z\right) ~,\eqno(46)$$

\noindent{where in these equations a prime denotes differentiation with
respect to $r$. As expected for any two--dimensional black hole, as described
above, we find that these form a system of coupled, linear, ordinary
differential equations. As noted in the Introduction, this model of
two--dimensional dilaton gravity does not incorporate propagating degrees of
freedom. This obviously does not imply that the linearized equations of motion
may not
be reduced to differentially--decoupled form. In order to proceed we would
like to attempt to follow the
prescription outlined in the previous section. Thus we must search for a
suitable transformation which will put the system into
differentially decoupled form, after which the final reduction to a
completely decoupled set of equations would proceed
without difficulty. In this connection we note that eq.(46) fixes the relation
between $\y$ and
$\z$, as a consequence of which we would naively expect a final reduction to
two decoupled second--order equations for the physical perturbation
functions. However, we also see that the distinct field perturbations appear
on an unequal footing in these equations: neither $\x$ nor $\y{''}$ appear in
these equations and, as we shall discover, this fact portends unusual
consequences.}

\noindent{We will begin the attempt to differentially decouple the
system given in eqs. (43) through (46) by noting that, in virtue of the
so--called Curci--Paffuti equations [26]}

$${1\over 2}\nabla_\n\b^{(\Phi)}=\nabla^\m\b_{\m\n}^{(g)}-2\b_{\m\n}^{(
g)}\nabla^\m\Phi ~,\eqno(47)$$

\noindent{where $\b^{(\Phi)}$ and $\b_{\m\n}^{(g)}$ are the
beta--functions{\footnote {$\b_{\m\n}^{(g)}$ and $\b^{(\F)}$ are
beta--functions of the non--linear sigma model with target
space metric given by eqs. (38) and (39), and generate the equations of
motion for
the $SL(2,R)/U(1)$ black hole.}} for the dilaton and gravitational fields,
respectively, we are guaranteed that
any single one of the equations of motion of the background fields is
automatically satisfied if the beta--functions corresponding to the remaining
equations vanish. This in turn
allows us to proceed to attempt to decouple the system of perturbation
equations by considering first eqs. (43) through (45),
without imposing eq.(46). To that end we can
first eliminate $\z{''}$ between eqs. (44) and (45) to obtain:}

$$\eqalignno{\x{''}&+2\x{'}(\sech~\csch-\th )-\z{'}(\sech~\csch+4\th )\cr
&-\o^2\z\chs-\sech~\csch~\y{'}+\y (\o^2\chs +8)=0 ~.&(48)\cr}$$

\noindent{We now write eqs. (43) and (48) as a simultaneous system:}

$${\cal M}X=Y ~,\eqno(49)$$

\noindent{where}

$${\cal M}=\pmatrix{\o^2\chs &-\sech~\csch\cr \o^2\chs+8&-\sech~\csch
\cr} ~,\eqno(50)$$
$$X=\pmatrix{\y\cr \y{'}\cr} ~,\eqno(51)$$
$$Y=\pmatrix{-\x{''}-2\coth~\x{'}-\sech~\csch~\z{'}-\o^2\chs\z\cr
-\x{''}-2\x{'}\left(\sech~\csch-\th\right)+\z{'}\left(\sech~\csch+4\th
\right)+\o^2\chs\z\cr} ~.\eqno(52)$$

\noindent{Using}

$${\cal M}^{-1}={1\over 8}\pmatrix{-1&1\cr
-\sinh r\cosh r\left(\o^2\chs+8\right)&\o^2\ch~\cosh^2r\cr} ~,\eqno(53)$$

\noindent{we find:}

$$4\y=2\tanh r~\x{'}+\left(\sech~\csch+2\th\right)\z{'}+
\o^2\chs\z ~,\eqno(54)$$

\noindent{and}

$$\eqalignno{4&\sech~\csch~\y{'}=\cr &=4\x{''}+2\left(\o^2+4\right)
\ch\x{'}\cr &~~~+\left[\o^2\chs\left(\sech~\csch+
2\th\right)+4\sech~\csch\right]\z{'}\cr &~~~+\o^2\chs\left(\o^2\chs+4\right)
\z ~.&(55)\cr}$$

\noindent{Differentiating eq.(54) we get:}

$$\eqalignno{4\y{'}=&2\tanh r~\x{''}+2\sss\x{'}
+\left(\sech~\csch+2\th\right)\z{''}\cr &-\left(\sss\css-\o^2\chs\right)\z{'}
-2\o^2\ch~\css\z ~,&(56)\cr}$$

\noindent{and combining eqs. (55) and (56) we obtain:}

$$\eqalignno{2&~\!\th\left(\sinh^2r+\cosh^2r\right)\x{''}+2\cosh^2r\left(\o^2
+4-\ssss\right)\x{'}\cr &-\left(\sech~\csch+2\th\right)\z{''}+\left(4+
\sss\css+2\o^2\cosh^2r\right)\z{'}\cr &+\o^2\ch{\Big [}\left(\omega
\cosh r~\!\ch\right)^2+2\left(2\cosh^2r+\css\right){\Big ]}\z=0 ~.&(57)\cr}$$

\noindent{Returning now to the basic perturbation equations (eqs. (43)
through (46)), we substitute the values for $\y$ and $\y{'}$ dictated by
eq.(54) into eq.(45), which, after some manipulation, yields}

$$\eqalignno{&\sss\z{''}-{\Big [}\o^2\ch+\qsech~\csch\left(\sinh^2r+
\cosh^2r\right){\Big ]}\z{'}-2\o^2\z\cr &~\!~~~~~~~~~~~~~-2\tanh^2r\x{''}-2
\tanh r\left(2+\sss\right)\x{'}=0 ~.&(58)\cr}$$

\noindent{Upon eliminating $\x{''}$ between eqs. (57) and (58) we find:}

$$-\chs\z{'}+\qch\left(\o^2\cosh^2 r
+2\right)\z+2\cosh^2 r\x{'}=0 ~.\eqno(59)$$

\noindent{Given this reduction, it is evident that the original system of
equations is not amenable to further reduction to differentially--decoupled
form through the use of a transformation of the dependent variables.
The most general simultaneous {\it linear}
transformation{\footnote{A nonlinear or more complicated non--local
transformation would be inconsistent with the restriction to small
fluctuations we have imposed throughout the analysis.}} of all of
the dependent variables when substituted into the set of equations
given by (43) through (45) fails to differentially decouple the system, which
at first appears to be surprising.{\footnote{The algebraic
manipulations required in the analysis of the general case are very involved.
A machine symbolic manipulation program, such as {\it Mathematica}, proves
extremely useful in sorting out the many pieces.}} That this is the case,
however, becomes apparent when one notices that eq.(46) is in fact
equivalent to eqs. (54) and (59). Thus, one may check that the original
system of perturbation equations given in eqs.(43) through (46) is actually
{\it entirely a consequence} of the following two first--order differential
equations:}

$$0=2\x{'}-\css~\z{'}+\cosh r~{\rm csch}^3r~\left(\o^2\cosh^2r+2\right)\z
{}~,\eqno(60)$$
$$0=2\y-\ch~\z{'}+\css~\z ~.\eqno(61)$$

\noindent{We therefore find that, for a given frequency $\o$, the
spatial evolution of the small fluctuations of the Witten solution is
completely determined by eqs. (60) and (61). We note that these are two
equations in three unknowns, and that only $\z$ and its derivative appear in
both eqs. (60) and (61). The consequence of this
is that one finds that a consistent first--order{\footnote {Technical
details involving boundary conditions on the fields and residual gauge
invariance,
respectively, are discussed in Notes 2 and 3 below, following the conclusion
section.}} perturbation solution may
be found for {\it any} choice of functional form for $\z$, which is a
completely unprecedented result. This result is altogether
different from the corresponding results one finds for any of the solved
small--fluctuation problems involving the known
black hole solutions in the general theory of relativity, where to date one
has always encountered eigenvalue equations ({\it cf} eq.(19)) for {\it some}
perturbation potential, and where in virtue of
the decoupling constraints it is {\it not} possible to find solutions for
arbitrary perturbations.{\footnote{One also
finds results similar to those which obtain in general relativity in the
case of the known ``string--inspired" solutions in four dimensions, {\it
i.e.}, for dilaton gravity in four dimensions [27].}} In the
usual case, the classical linear response of the black hole may
be determined once the perturbation potentials are known. The linear
response is defined by the scattering coefficients, which, for an assumed
asymptotic behavior, are uniquely predicted by the radial eigenvalue
equations. In the present case, in contrast, the equations admit a continuous
infinity of solutions. Within this set of solutions are entirely distinct
functions with
identical asymptotic behavior, and thus the equations do not uniquely
determine scattering coefficients. Therefore, although eqs. (60) and (61) may
be unambiguously solved, it is nevertheless impossible to unambiguously
ascertain the linear response of the black hole.
We may thus say that the equations which normally determine the linear
response of the black hole are in this case {\it physically unpredictive.}
{\footnote {Note that
this is not the same as the condition of a vanishing perturbation
potential, $v_i=0$, for which one would have a reduced system of the
form $\d p_i{''}+\o^2\d p_i=0$.}}}

\noindent{We may gain perspective on the surprising behavior of the
$SL(2,R)/U(1)$ black hole which we have discovered by viewing our results
in the context of the underlying conformal field theory. In particular
we shall consider the dimension $(1,1)$ operators of the conformal
field theory: the so--called marginal operators. These operators have the
property that when one or more of them is incorporated into the definition of
the sigma model the value of
the central charge is preserved. There is a {\it subset} of these operators
which are further distinguished by the property that the conformal dimensions
of all operators in this subset are preserved as well in the modified model.
This special class of marginal operators are known as exactly marginal
operators. Their properties have been elaborated in [8], where it was shown
how to explictly compute them to first--order in an expansion in
$1/k$.{\footnote {In the limit that $k\to\infty$ one obtains purely
{\it classical} conformal field theory.}} The spacetime effect on the
background fields of specific exactly marginal operators (evaluated to
first--order in $1/k$) was derived in
[8]. We shall now consider the compatibilty of this action with the
conditions embodied within the basic perturbation equations of the
$SL(2,R)/U(1)$ black hole.
The particular exactly marginal operators investigated in [8] were
the operators $L_0^1{\bar L}_0^1$ and $L_0^2{\bar L}_0^2$.
{\footnote {Here
$L_n^s$ is defined as $L_n^s=V_n^s+{\tilde V}_n^s$, where $V_n^s$ and
${\tilde V}_n^s$ are the
$n$'th Fourier components of two of the generators of the super--$W_\infty$
algebra, and $s$ is the $W_\infty$ ``spin" of the algebra [8].}}
It was shown in [8] that the addition of the operator
$L_0^2{\bar L}_0^2$ to the action of the non--linear sigma model generates
the deformed lagrangian $L$ given by ($\a$ is an arbitrary parameter)}

$$L=\partial_zr\partial_{\bar z}r{\Bigg [}1-2\a\left(\css+\sss\right){\Bigg ]}
+\partial_z\theta\partial_{\bar z}\theta{\Bigg [}\sinh^2r+2\a-{\left(\sinh^2r
+2\a\right)^2\over \cosh^2r+2\a}{\Bigg ]} ~.\eqno(62)$$

\noindent{We would like to determine whether or not this deformation, produced
by an exactly marginal operator, is encompassed within the continuous infinity
of allowed deformations we have discovered in our analysis of the small
fluctuations of the black hole.{\footnote {Note that it is appropriate to
ask this question since the deformation in eq.(62) has been computed to
lowest--order in an expansion in $1/k$, which is to say that it represents
a classical conformal field theoretic effect. As such, it is consistent to
compare it with our analysis of the small fluctuations since it has also been
(implicitly) performed at lowest order in $1/k$.}} In comparing the
deformation produced by the operator $L_0^2{\bar L}_0^2$ with our analysis of
the small fluctuations it is important to note that the calculation in [8]
leading to eq.(62) was performed with the neglect of terms in the sigma model
which were of higher than bilinear order in derivatives. With the proviso that
the metric is asymptotically--flat it is legitimate to neglect these terms in
the limit $r\to\infty$. Thus, we may read off from eq.(62) the appropriate
fields to substitute into the perturbation equations given in eqs. (60)
and (61), taking care to work in the large--$r$ limit. It is
straightforward to
verify that eqs. (60) and (61) are indeed satisfied in this limit, and
we thus find that {\it a particular} example of a first--order fluctuation
which is consistent with the linear constraints given by the basic
perturbation equations is provided by the operator $L_0^2{\bar L}_0^2$.
However, the analogous calculation applied to
the operator $L_0^1{\bar L}_0^1$ reveals that the deformation
it generates corresponds to the excitation of a {\it non--linear} departure
from the background [28]. Specifically, the tachyon field, which is implicitly
present in the background with zero field strength in the Witten solution,
appears as a second--order perturbation. However, we have restricted our
analysis to {\it small} fluctuations understood to be of first--order,
and it is thus inappropriate to compare the effect of this operator with our
results. The complete set of all exactly marginal operators is believed to
constitute a {\it countably} infinite set, since the quantum numbers which
distinguish them are discretely valued. Clearly these cannot encompass all
of the allowed deformations we have discovered, since, as we have demonstrated
above, the linearized perturbation equations of the $SL(2,R)/U(1)$ black hole
allow a {\it continuous} infinity of solutions. Although we have
demonstrated that the operator $L_0^2{\bar L}_0^2$ at
large $r$ generates
a particular one of the continuous infinity of deformations we have
discovered, the fact that the operator $L_0^1{\bar L}_0^1$
does {\it not} generate a small fluctuation suggests that only some (and
perhaps none) of the remaining exactly marginal operators excite small
fluctuations.{\footnote {Actually, the fact that we have shown that the
exactly marginal operator $L_0^2{\bar L}_0^2$ excites an allowed small
fluctuation serves to verify the {\it consistency}, to first--order in
$1/k$, of the two $1/k$ expansions, used to derive the black hole and to
explicitly compute $L_0^2{\bar L}_0^2$, respectively.}} Thus the mere
existence of an infinite set of exactly marginal operators does not imply
that there are an infinity of allowed small fluctuations, and even if it did,
this would have accounted for only a countable infinity. We have thus
discovered a new, continuously infinite class of motions the fundamental
origin of which awaits explanation.}

\noindent{One may therefore enquire as to precisely where the new continuous
infinity of allowed small fluctuations we have discovered fits in the
description of the physics of the $SL(2,R)/U(1)$ black hole. The (subset of
the) countably infinite set of exactly marginal operators consistent with the
basic perturbation equations is evidently insufficient to describe all of
the allowed motions of the black hole. This black hole is actually a
particular two--dimensional solution to the equations of motion of string
theory. More precisely, the Witten black hole is an {\it approximate} solution
to the string equations of motion: It is a solution at the level of the Born
approximation in string theory since the sigma model has been formulated on a
sphere and thus all higher--loop (and, more generally, non--perturbative)
string corrections have been ignored; It is evidently an approximate solution
to the sigma model as well, as reflected in the presence of $O(1/k^2)$
corrections to the value of the central charge pointed out by Witten in [2].
It is natural to speculate that what is missing from the picture lies in the
corrections that have been neglected in the higher string--loop contributions,
or in the higher--order $1/k$ contributions on the sphere, or perhaps some
combination of both contributions. The underlying $W_\infty$ structure of
this model appears to be related to the existence of an infinite number of
exactly marginal operators. However, as we have stressed, it is not obvious
that these operators generate a (countably infinite) set of small
fluctuations, and
we expect this to remain true even if one were to consider the effect of the
exactly marginal operators computed to {\it all} orders in $1/k$. As we have
also stressed, however, even if the exactly marginal operators computed to
all orders in $1/k$ {\it did} excite a countably infinite set of modes, this
would not
account for the continuous infinity of perturbations we have found. Ideally,
one would like to compare the effects of exactly marginal operators,
calculated to all orders in $1/k$, to small fluctuations as determined by
the {\it exact} beta--functions. The latter, unfortunately, are not known
at present,
although the solution to the equations they correspond to
(with the same leading order behavior as the Witten
black hole) has been calculated [6,9]. It should in any event be worthwhile
to extend our results by examining the next--to--leading--order corrections
in $1/k$.}

\sect{{\bf II.2.c Massive Dilaton Black Holes}}

\noindent{We will now study the small fluctuations of two--dimensional
black hole configurations in which the dilaton is massive, and we thus return
to eqs. (25) and (26). In order to proceed it is necessary to
select a particular form for the potential energy density $V\left(\Phi
\right)$. Here one has a great deal of latitude since, apart from a special
case such as the Witten solution ({\it i.e.}, choosing $\tilde V={\tilde V}'
=0$) which furnishes a solution at the
level of the Born approximation to the equations of motion of string theory,
the models defined by the action of eq.(22) are no more than
``string--inspired" models. Thus, the fact that
it is not today known how (or better, if) string--theoretic principles
determine the form of the dilaton potential is to a certain extent
unimportant. We shall here follow the choice made in recent studies of these
configurations [19] in which the potential is chosen by {\it fiat} to be of
the form:}

$$V\left(\Phi\right)=m^2\Phi^2 ~,\eqno(63)$$

\noindent{where $m$ is the mass of the dilaton.{\footnote {It must be
remembered {\it throughout} the following analysis that, as stated above in
the text and in footnote \#7, in
our calculations we make the substitution $\F\to -\F/2$ in the equations of
motion. This should be borne
in mind when comparing certain expressions below with corresponding
expressions given in [19]. In particular with the choice of $V$ given in
eq.(63), one has $\tilde V={1\over 4}m^2\F^2$.}} This is certainly the
simplest non--trivial choice one may make for the potential, and it is
conceivable that such a choice may prove to be useful. Upon
substituting eq.(63) into eqs. (25) and (26) one then obtains:}

$$0=\nabla^2\Phi+\left(\nabla\Phi\right)^2-4\L^2+{1\over 4}m^2e^\Phi
\Phi^2 ~,\eqno(64)$$

$$R_{\m\n}=\nabla_\m\nabla_\n\Phi+{1\over 2}m^2e^\Phi g_{\m\n}\left(
{1\over 4}\Phi^2+{1\over 2}\Phi\right) ~.\eqno(65)$$

\noindent{In recent studies a putative massive dilaton black hole
configuration was studied by employing the ansatz of eq.(20) for the metric
tensor, with the metric functions taking the values}

$$g_{00}=-e^{2f_0}=-A^2,~~~~~~~~g_{11}=e^{2f_1}=A^{-2} ~,\eqno(66)$$

\noindent{where $A=A\left(r\right)$ is to be
determined by solving the field equations. It was shown that there
exist two possible black hole solutions: one for which
the dilaton field strength is given by a constant: $\Phi=p_0$, say, and
another for which the dilaton field is proportional to $r$: $\Phi=
p_1r$.}

\sect{{\bf II.2.c.i Constant Dilaton Solution}}

\noindent{In the case of a constant dilaton field, $\Phi=p_0$, one may
prove that the constant scalar curvature is given by{\footnote {See the
comment in footnote \#18.}}}

$$R=-\left(A^2\right)''=4\L^2\left(1+2p_0^{-1}\right) ~,\eqno(67)$$

\noindent{as a result of which we find the metric solution $A^2=ar^2+
br+c$, where}

$$a=-2\L^2\left(1+2p_0^{-1}\right) ~,\eqno(68)$$

\noindent{and $b$ and $c$ are integration constants. We will now
consider
the basic perturbation equations (eqs. (30) through (31)) for the constant
dilaton solution. We observe first that the $(01)$--component of the
linearized Einstein equations is given by}

$$0=i\o\left(\z{'}-A^{-1}A'\z\right) ~,\eqno(69)$$

\noindent{which immediately yields the integral $\z=\k A$ with $\k$ a
constant. We must now ensure that our small--fluctuation
approximation is valid, which is the case if $|\z/\F|\ll 1$. For the
constant dilaton solution this means that we must have}

$$|\z(r)/\F|={\Bigg \vert}{\k\over p_0}{\sqrt {ar^2+br+c}}{\Bigg \vert}
\ll 1 ~.\eqno(70)$$

\noindent{We will now prove that this inequality dictates that we must
take $\k=0$ for the value of the integration constant. The smallness
constraint must be satisfied everywhere in order to justify the neglect
of terms of higher than first--order in our analysis, and in particular in
the limit $r\to\infty$. From eq.(70) we see that we must have}

$$\lim_{r\to\infty}|\z/\F|=\lim_{r\to\infty}{\Bigg \vert}{\k a^{1/2}\over
p_0}r{\Bigg \vert}\ll 1 ~,\eqno(71)$$

\noindent{which implies that $\k=0$, or that $a$=0, or both.{\footnote
{From eq.(68) we see that $a$ can vanish for special values of $p_0$ or
$\L$.}} However, if $a=0$ we have}

$$\lim_{r\to\infty}|\z/\F|=\lim_{r\to\infty}{\Bigg \vert}{\k b^{1/2}\over p_0}
r^{1/2}{\Bigg \vert}\ll 1 ~,\eqno(72)$$

\noindent{which implies that $\k=0$, or that the integration constant
$b=0$, or both. However, if $a$=$b$=0, one has $A^2=c$, in which case
for arbitrary non--vanishing{\footnote {In the special case $a=b=c=0$ the
metric function $A^2$ vanishes identically in which case the metric tensor
is ill--defined globally. In any event, we note in passing that for this
case one obtains eq.(73) automatically.}} $c$ the metric tensor is constant
and non--singular ({\it cf} eq.(66)), and the configuration is no
longer a black hole at all.{\footnote {Note that this argument is distinct
from the observation of the fact that when $a=0$ the curvature vanishes ({\it
cf} eq.(67)). The constant
dilaton configuration is a black hole in virtue of the fact that there is an
event horizon, and {\it not} because there is a curvature singularity, as
indeed there is not.}} Therefore we must require that $\k=0$, as a result of
which we have found that}

$$\z=0 ~,\eqno(73)$$

\noindent{and thus all first--order fluctuations in the dilaton field have
exactly vanishing amplitude. As a result of this we observe that the
linearized dilaton equation ({\it cf} eq.(30)) vanishes identically.
Furthermore, making use of eq.(31) we find that the $(00)$-- and
$(11)$--components of the Einstein equations simplify dramatically, and
we obtain}

$$\eqalignno{&A^4{\Big [}\x{''}+3A^{-1}A'\x{'}-A^{-1}A'\y{'}+2\left(\x
-\y\right)\left(A^{-1}A{''}+A^{-2}A{'}^2\right){\Big ]}+\o^2\y\cr
&~~~~~~~~~~=-{1\over 2}A^2m^2e^{p_0}\left(p_0+{1\over 2}p_0^2
\right)\x ~,&(74)\cr}$$

\noindent{for the $(00)$--equation, and}

$$-\x{''}-3A^{-1}A'\x{'}+A^{-1}A'\y{'}-\o^2A^{-4}\y={1\over 2}A^{-2}m^2
e^{p_0}\left(p_0+{1\over 2}p_0^2\right)\y ~,\eqno(75)$$

\noindent{for the $(11)$--equation. Upon multiplying the $(11)$--equation
by $A^4$ and adding the result to the $(00)$--equation one obtains}

$$\left(\x-\y\right){\Bigg [}A^3A''+A^2A{'}^2+{1\over 4}A^2m^2e^{p_0}
\left(p_0+{1\over 2}p_0^2\right){\Bigg ]}=0 ~.\eqno(76)$$

\noindent{The equations of motion of the background fields may now be used to
obtain the relation $16\L^2=m^2p_0^2e^{p_0}$. Upon substituting this
expression
into eq.(76), along with the value of $A^2$ with $a$ given by eq.(68),
one finds that the quantity in the square brackets vanishes identically,
and thus the two gravitational equations form a redundant system and there is
only one independent equation.
The consequence of this is that the black hole coupled to a massive dilaton
with constant field strength behaves in a manner similar to that of the
$SL(2,R)/U(1)$ black hole [2]: a solution for one of the gravitational
perturbations may be found for {\it any} choice of the other one. As before,
we must ensure that the solutions are sufficiently small to be considered
as first--order perturbations. Since we have shown that eqs. (74) and (75)
are equivalent, we may check that the fluctuations are acceptable by
considering either one of them. To that end we note that the $(11)$--equation
may be written as}

$$0=A^4\x{''}+3A^3A'\x{'}-A^3A'\y{'}+\o^2\y+\varpi A^2\y ~,\eqno(77)$$

\noindent{where $\varpi={1\over 2}m^2e^{p_0}\left(p_0+{1\over 2}p_0^2
\right)$ is a constant. Eq.(77) can be rewritten as}

$$0=A\left(A^3\x{'}\right)'-{1\over 4}\left(A^4\right)'\y{'}+\left(
\o^2+\varpi A^2\right)\y ~,\eqno(78)$$

\noindent{which may be integrated to yield}

$$A^3\x{'}=\psi+\int_{r_h}^r dr~{\Bigg [}{1\over 3}\left(A^3\right)'\y{'}
-\left(\o^2+\varpi A^2\right)A^{-1}\y{\Bigg ]} ~,\eqno(79)$$

\noindent{where $r_h$ is the position of the event
horizon and $\psi$ is the constant of
integration. Now, since $A^2=ar^2+br+c$, near the horizon one has}

$$A\sim\left(r-r_h\right)^{1/2},~~~~~A^3\sim\left(r-r_h\right)^{3/2}
{}~,\eqno(80)$$

\noindent{{\it etc}. Noting that $\x{'}=A^{-1}A'$, we will find it
convenient to ensure that $|\x/f_0|\ll 1$ by proving the sufficient
condition that $|\x{'}/f_0'|\ll 1$. Substituting eq.(80) into eq.(79) we
find}

$$\left(r-r_h\right)^{-3/2}{\Big \{}\psi+\left(r-r_h\right)^{1/2}\x-
\int_{r_h}^r dr~{\Big [}\o^2\left(r-r_h\right)^{-1/2}+\varpi\left(r-r_h
\right)^{1/2}{\Big ]}\y{\Big \}}\ll\left(r-r_h\right)^{-1} ~.\eqno(81)$$

\noindent{This condition requires that we take $\psi=0$ for the value of the
integration constant, as a result of which the constraint will be satisfied
as long as $\y$ is regular in the limit $r\to r_h$. One can similarly check
that a continuous distribution of small gravitational fluctuations can be
found in the limit $r\to\infty$. Thus, as is the case for the Witten black
hole, there exist a continuous infinity of small--fluctuation solutions to the
linearized equations of motion for the massive dilaton black hole with
constant field strength, and it is therefore {\it in principle} impossible to
unambiguously determine the classical linear response of the black hole.
Of course, unlike the $SL(2,R)/U(1)$ black hole, here the
fluctuation in the dilaton field is constrained to vanish, but there remains
an uncountably infinite ambiguity in the gravitational perturbations. Although
this black hole is characterized by a massive dilaton and is therefore not
described in terms of a conformal field theory, whereas the $SL(2,R)/U(1)$
black hole is so described, the two different two--dimensional configurations
display similar behavior: the classical linear response is indeterminate, an
unusual situation which differs radically from the behavior of all
known black holes in four dimensions.}

\sect{{\bf II.2.c.ii Linear Dilaton Solution}}

\noindent{In the case of a linear dilaton solution with $\Phi=p_1r$
the equations of motion for the background fields have been solved by
Gregory and Harvey [19], who find the following expression for the metric
function $A$:{\footnote {See the comment in footnote \#18.}}}

$$A^2=1-2Me^{\pm p_1r}-{m^2\over 16p_1^2}e^{\mp p_1r}\left(2p_1^2r^2
\pm 2p_1r+1\right) ~,\eqno(82)$$

\noindent{with $M$ (not to be confused with the dilaton mass $m$) arbitrary.
With the background metric specified by
eq.(66) we may consider the basic perturbation equations for the black
hole. We obtain}

$$0=i\o\left(\z{'}-A^{-1}A'\z-p_1\y\right) ~,\eqno(83)$$

\noindent{from the $(01)$--component of the linearized Einstein equation, and}

$$\eqalignno{0=&A^2\z{''}+{\Big [}\left(A^2\right){'}+2p_1A^2{\Big ]}\z{'}+
{\Big [}\o A^{-2}+{1\over 4}p_1m^2re^{p_1r}\left(2+p_1r\right){\Big ]}\z\cr
&+p_1A^2\x{'}-2p_1{\Big [}\left(A^2\right){'}+p_1A^2{\Big ]}\y-p_1A^2
\y{'} ~,&(84)\cr}$$

\noindent{for the linearized dilaton equation. For the $(00)$--component of
the linearized Einstein equations we obtain}

$$\eqalignno{A^4&{\Big [}\x{''}+3A^{-1}A'\x{'}-A^{-1}A'\y{'}+2\left(\x-\y
\right)\left(A^{-1}A''+A^{-2}A{'}^2\right){\Big ]}+\o^2\y\cr &=-\o^2\z-{1
\over 2}
A^2\left(A^2\right)'\z{'}+p_1A^2{\Big [}\left(A^2\right)'\y-\left(A^2\right)'
\x-A^2\x{'}{\Big ]}\cr &~~~~-{1\over 2}e^\Phi{\Bigg [}A^2m^2\left(\Phi+{1\over
2}\Phi^2\right)\x+A^2m^2\left({1\over 4}\Phi^2+\Phi+{1\over 2}\right)\z
{\Bigg ]} ~,&(85)\cr}$$

\noindent{and we find}

$$\eqalignno{&-\x{''}-3A^{-1}A'\x{'}+A^{-1}A'\y{'}-\o^2A^{-4}\y\cr
&~~~~~~~~=\z{''}-p_1\y{'}+A^{-1}A'\z{'} ~&(86)\cr}$$

\noindent{for the $(11)$--component of the linearized Einstein equations.
By appropriately combining these equations and making use of the
expression for $A(r)$ given in eq.(82), we
may rewrite the system as{\footnote{We have taken $M=0$, to ensure an
asymptotically--flat metric, as well as the upper choice of sign in
eq.(82).}}}

$$0=\a(r)\z{'}+\b(r)\z ~,\eqno(87)$$
$$0=\g(r)\x+\e(r)\z{'}+\r(r)\z ~,\eqno(88)$$
$$0=-p_1\x{'}+\t(r)\z{'}+\s(r)\z ~.\eqno(89)$$

\noindent{In these equations
the scalar functions $\a(r)$ and  $\b(r)$ are given by{\footnote
{The expressions
for $\g$, $\e$, $\rho$, $\t$ and $\s$ are huge and will not be displayed
here. A {\it Mathematica} routine which generates these functions will be
provided via electronic mail upon request.}}}

$$\a(r)\equiv -8e^{3p_1 r}m^2r (2+p_1 r) ~,\eqno(90)$$

$$\eqalignno{\b(r)\equiv &I^{-1}(7m^4-832e^{2p_1r}m^2p_1^2+40m^4p_1r+
48e^{3p_1r}m^4p_1r+
1792e^{2p_1r}m^2p_1^3r+192m^4p_1^2r^2\cr &+216e^{3p_1r}m^4p_1^2r^2-
512 e^{2p_1r}m^2p_1^4r^2-576 m^4p_1^3r^3-288e^{3p_1r}m^4p_1^3r^3+
192m^4p_1^4r^4\cr &-192e^{3p_1r}m^4p_1^4r^4) ~,&(91)\cr}$$

\noindent{where}

$$I\equiv m^2-64e^{2p_1r}p_1^2+4m^2p_1r+8m^2p_1^2r^2 ~.\eqno(92)$$

\noindent{The analysis of this coupled
system of differential equations proceeds as follows. One first differentiates
eq.(88), which may be used to eliminate all terms proportional to both $\x$
and $\x{'}$ across eqs. (88) and (89), and hence from the complete system
since no such terms appear in eq.(87). Then eq.(87) and its derivative may
be used to eliminate all terms proportional to $\z{'}$ and $\z{''}$ from the
system as well. The result of these successive operations is a single equation
of the form{\footnote{The explicit form of the function
$\chi(r)$ is extremely complicated and will not be given here. A
{\it Mathematica}
routine which generates this function will be provided via electronic mail
upon request. The interested reader is warned that the output file is
exceedingly large, consuming approximately 100 kilobytes of computer
memory.}}}

$$\chi(r)\z(r)=0 ~.\eqno(93)$$

\noindent{The next step in the
analysis entails a numerical examination
of the function $\chi(r)$, which demonstrates that in general one has
$\chi(r)\not= 0$, as may be seen in Table 1 where representative values of
$\chi(r)$ are displayed. This result suggests that $\z=0$. One may then
also note that eq.(87) can be directly integrated to yield}

$$\z={\rm const}.\exp\left(-\int^r dr \b/\a\right) ~.\eqno(94)$$

\noindent{Given that generically $\chi(r)\not= 0$, and that this is true
in particular for values of $r$ for which $\b/\a$ is finite (as may easily be
checked), the above equation can be consistent
with the remaining equations ({\it i.e.}, eqs. (88) and (89), or, what is the
same thing, with eq.(93)) only
if the integration constant vanishes identically. Inspection of eq.(83)
reveals that one must take $\y=0$ for consistency. Finally, a numerical
analysis of the function $\g\left(r\right)$ demonstrates that in general one
has $\g\left(r\right)\not= 0$, as may be seen in Table 2, where representative
values of $\g(r)$ are displayed, in virtue of which one must take $\x=0$
for consistency ({\it cf} eq.(88)). This analysis demonstrates
quite generally that the only consistent simultaneous solution of the
coupled system of perturbation equations is the trivial solution in which
{\it all} of the small fluctuations are constrained to vanish. Thus, we have
found another unexpected result: the linear dilaton species of
two--dimensional massive
dilaton black hole does not admit {\it any} small fluctuations around the
background configuration, in complete contrast once again to the corresponding
results which have been obtained for the black holes of four--dimensional
general relativity. The result
indicates that the black hole coupled to a massive, linear dilaton
represents an isolated point in the space of field configurations of
two--dimensional dilaton gravity.}

\sect{{\bf III. Conclusions}}

\noindent{We have found that the Witten black hole behaves in a radically
different
way from all other known black hole solutions, whether in the conventional
general theory of relativity or in four--dimensional dilaton gravity. For
those solutions one may perform an analysis (as outlined in Section II.1
above)
of the linear response of the black hole to incoming waves which leads to
decoupled eigenvalue equations for the physical fluctuations characterized
by specific perturbation potentials. For these various black holes
one finds different perturbation potentials corresponding to
different varieties of uniquely determined scattering behavior, and indicative
of whether or not a bound state can form. In the case of the Witten solution,
however, the equations for the small fluctuations cannot be brought into
completely decoupled form. In contrast to the
situation which obtains for all previously studied black holes, there exist a
continuous infinity of acceptable ({\it i.e.}, sufficiently small to
be considered of first--order) solutions to the linearized equations of motion
about the background. We have further shown that as a consequence of this
it is impossible to unambiguously determine the
classical linear response of the black hole, since the reduced perturbation
equations do not uniquely determine the scattering coefficients
for specified asymptotic behavior.}

\noindent{In studying a two--dimensional conformal field theory it
is interesting to study the exactly marginal $(1,1)$ operators. In the case
of the conformal field theory underlying the $SL(2,R)/U(1)$ black hole some of
the exactly marginal operators may generate deformations
of the action of the underlying sigma model which correspond to small
fluctuations of the background fields of the black hole. We have explicitly
confirmed this for
the particular case of the exactly marginal operator $L_0^2{\bar L}_0^2$ by
verifying that the small fluctuations it produces do indeed satisfy the
linearized equations of motion. However, the exactly marginal operators
constitute only a countably infinite set, and in any event, as we
have discussed, only {\it some} of them will excite physically--acceptable
small fluctuations. Thus it is necessary to look elsewhere in order to account
for the complete, uncountably infinite set of small motions which our
equations
allow the black hole to perform. It is very surprising to encounter such an
intrinsic ambiguity in the classical analysis of the linear response. However,
we may recall that the
$SL(2,R)/U(1)$ black hole is a solution at the level of the Born approximation
to the equations of motion of a string propagating in two
dimensions. The black hole configuration is approximate as well in that
higher--order corrections in an expansion in powers of $1/k$ are neglected in
obtaining the solution. Although this approximate character is well--known,
the hope has been expressed by many authors that the black hole solution is
nevertheless ``very useful for getting a qualitative picture of the physics."
We suspect that the behavior we have uncovered, which is highly unusual,
is sufficiently different
from the behavior of all known four--dimensional black holes that it may
be misleading to utilize this black hole
model at all as a point of reference in studying the properties of
physically--realistic black holes in four dimensions. It is natural to wonder
whether a proper classical linear response can be restored by considering
instead a black hole solution which is {\it exact}. Of course, the word exact
has a double meaning here. One's chief desire would
be to have in hand a black hole solution which is truly exact in the sense
of string theory, in which all string--loop corrections have been accounted
for. Such a solution is not available at the moment, and may not be known for
a long, long time. On the other hand, when considered solely as a black hole
{\it qua} a solution to a two--dimensional theory of gravitation, one might
hope that a proper
linear response would be obtained by analyzing instead the
corresponding
two--dimensional black hole solution in which higher--order $1/k$
corrections on the sphere have been included. Dijkgraaf, {\it et. al.} [6] and
Bars and Sfetsos [9]
have claimed to have derived such a solution, and work is
in progress in extending the analysis of this paper to that black hole.}

\noindent{As discussed in
Section I.2, there are a number of related black hole constructions which have
been discovered recently. In particular, we examined two specific examples of
related black hole solutions which have been found. These are both
two--dimensional black holes coupled to a massive dilaton. In the somewhat
special case in which the background dilaton is characterized by a {\it
constant} field strength, we find behavior reminiscent of the $SL(2,R)/U(1)$
black hole, in that a continuous infinity of small fluctuations is admitted
by the linearized equations of motion, and it is again impossible in
principle to ascertain the classical linear response of the black hole. That
this black hole behaves in a manner similar to the $SL(2,R)/U(1)$
black hole is surprising in that one might have thought that the essential
source of this unusual behavior in the case of the Witten solution might
lie in its origin as a conformal field theory. However, since the
constant dilaton black hole is in particular coupled to a massive dilaton, and
is thus not derived from a conformal field theory, that explanation is open to
question. We also analyzed the linear response of the two--dimensional black
hole solution coupled to a massive, {\it linear} dilaton. This is an important
example to consider since the linear dilaton vacuum is roughly analogous
to four--dimensional Minkowski space. Here we found entirely
different, but again unexpected, behavior as compared to the linear
response of known four--dimensional black holes. In striking contrast to the
other examples we studied, the black hole with massive linear dilaton
is intrinsically constrained so that no small fluctuations are allowed at
all. Thus this black hole configuration is an isolated point in the
space of field configurations of
the theory of two--dimensional dilaton gravity, and as such represents
an unusual occurrence in a generally covariant theory.}

\noindent{These surprising results do not appear to be a consequence of the
fact that the underlying dilaton
gravity theories do not incorporate propagating degrees of freedom. All of the
black holes we have studied share this property, yet they display two vastly
different linear response behaviors. In this connection we note that attention
has recently turned to the study of the CGHS black hole [3].
This black hole is being closely studied in an attempt
to resolve questions of four--dimensional black hole physics, such as: What is
the nature of the final result of the
Hawking radiation process? Do black holes destroy information? If they
do, does this signal that the very tenets of the quantum theory itself
must be modified? Thus, the obvious candidate two--dimensional black hole
which must,
and which remains to be, analyzed using the methods of this paper is the CGHS
solution. As we have discussed, the fundamental cause of the infinite
classical fluctuation ambiguity found for the $SL(2,R)/U(1)$ black
hole is not yet clear. Nevertheless, the CGHS model has its origin
in a non--linear
sigma model which is closely related to that which underlies the Witten black
hole. This suggests that the CGHS model may well also
display classical linear response behavior which is radically different
from that of all known four--dimensional black holes. Recall
that the problems for which the CGHS model is being studied
are inherently quantum mechanical in origin, and that
the correspondence principle dictates
that one must properly recover classical mechanics from quantum mechanics in
the appropriate limit. However, the two--dimensional black holes we have
analyzed in detail in this paper do not display
linear response behavior which is in any way characteristic of their
four--dimensional counterparts. If the classical mechanical behavior of
the CGHS black hole is indeed shown to be radically different from that
of four--dimensional black holes, then its use as a toy model
from which to draw inferences applicable to the outstanding problems of
four--dimensional black holes must be treated with caution.}

\noindent{{\it Acknowledgements:} The authors wish to thank S. Chaudhuri,
S. Giddings, T. Jacobson, J. Lykken and J. Sucher for comments.}

\sect{{\bf {\underbar {Notes}}}}

\Item{{\bf Note 1}}

\noindent{As the present paper was being completed the authors received
a preprint (reference [29] by Diamandis,
{\it et. al.}, in which related issues involving black holes with
time--dependent tachyons are treated.}

\Item{{\bf Note 2}}

\noindent{In order to ensure
consistency with our assumption
throughout that all fluctuations are small (and hence that only linear terms
need be retained) one must, of course, choose $\z$ such that $|\z/\F|\ll
1$. With the help of eq.(40) we see that this requires that}

$$|\z/\F|={\Bigg \vert}{\z\over\ln\cosh^2r+{1\over 2}\ln\left(k/2\right)
+\ln M}{\Bigg \vert}\ll 1 ~.\eqno({\rm N}2.1)$$

\noindent{Having chosen $\z$, as stated in the text, one may always find a
solution for the fluctuations in the metric by substituting it into
eqs. (60) and (61). We must restrict our attention, however, to solutions
for the metric perturbations which satisfy the constraints
$|\x/f_0|\ll 1$ and $|\y/f_1|\ll 1$. It is easy to see that there are
a continuous infinity of simultaneous solutions which satisfy these
constraints. For instance, by solving eq.(61) for $\y$ and using eq.(39)
we find that}

$$|\y/f_1|={\Bigg \vert}\ch~{\z{'}\over \ln\left(k/2\right)}-\css~{\z\over
\ln\left(k/2\right)}{\Bigg \vert} ~.\eqno({\rm N}2.2)$$

\noindent{We are interested only in the behavior of fields at points outside
of the event horizon, which is located at $r=0$. Thus it is clear that we
must consider the amplitudes of the field perturbations at the two extreme
locations: $r=0$ and $r\to\infty$, since it is only at these positions that
it is possible for the necessary ``smallness" constraints to be violated.
We find that we must choose
those perturbations in the dilaton field such that
$\z$ vanishes at infinity and goes to zero faster than $r$ at the
horizon. Similarly, we may solve eq.(60) for $\x{'}$ and then integrate both
sides of the resulting equation. After an integration by parts, and making
use as well of eq.(38), we obtain}

$$|\x/f_0|={\Bigg \vert}{{\rm csch}^2r~\z\over \ln\left[\left(k/2\right)
\tanh r\right]}-{\o^2\int_0^r dr~\qch~\z\over\ln\left[\left(k/2\right)
\tanh r\right]}{\Bigg \vert} ~.\eqno({\rm N}2.3)$$

\noindent{We see that in order to satisfy the condition $|\x/f_0|\ll 1$ it is
necessary
again to require that $\z$ vanish at infinity and
go to zero faster than $r$ at the event horizon. These simple
conditions can obviously be satisfied for a {\it continuous} distribution of
choices of values of the fluctuation $\z$. Having established that the
magnitudes of the ratios of the fluctuations to the background fields are
finite at the horizon and at
infinity we are done, since it follows from the above equations that
they are finite at all intermediate values of $r$. To see this, and in
particular to see that the ratios are both finite and small, recall that we
are performing a classical analysis,
which is to say that we are actually working in the limit $k\to\infty$. Thus,
we are assured that the smallness constraints are satisfied for all of the
fluctuations, in view of which we have confirmed
that there are an uncountable infinity of physically acceptable solutions
to the basic perturbation equations of the $SL(2,R)/U(1)$ black hole.}

\Item{{\bf Note 3}}

\noindent{In this note we discuss the residual gauge freedom implicit in
our construction, and its effect on the infinite set of solutions to the
linearized equations of motion. We have chosen ({\it cf} eq.(21)) the
following ansatz for the perturbed metric}

$$\pmatrix{-e^{2f_0}&0\cr 0&e^{2f_1}\cr}~~\rightarrow ~~\pmatrix{
-e^{2f_0+2\d f_0}&0\cr 0&e^{2f_1+2\d f_1}\cr} ~.\eqno({\rm N}3.1)$$

\noindent{In setting the off--diagonal components to zero one has
only partially fixed the gauge. Clearly, the ansatz in eq.(N3.1) is
unaffected by coordinate transformations of the form}

$$t\rightarrow {\tilde t}={\tilde t}(t),\ ~~~r\rightarrow {\tilde r}=
{\tilde r}(r)~,\eqno({\rm N}3.2)$$

\noindent{where ${\tilde t}$ and ${\tilde r}$ are arbitrary functions of $t$
and $r$, respectively.  However, the background should remain unchanged under
this transformation, and therefore we must have}

$${\tilde t}=t+g(t),\ ~~~{\tilde r}=r+h(r) ~,\eqno({\rm N}3.3)$$

\noindent{where $g(t)$ and $h(r)$ are of the same order of smallness
as the $\d f_i$.  Upon utilizing the standard transformation laws for the
metric tensor and for the dilaton, one finds}

$$\d f_0\rightarrow \d f_0+g'(t)+f_{0,t}g(t)+f_{0,r}h(r),\ ~~~\d f_1
\rightarrow\d f_1+h'(r)+f_{1,t}g(t)+f_{1,r}h(r)~,\eqno({\rm N}3.4)$$
$$\d\Phi \rightarrow \d\Phi+g(t)\Phi_{,t}+h(r)\Phi_{,r} ~,\eqno({\rm N}3.5)$$

\noindent{where a prime denotes differentiation with respect to the
argument. Recall that we require that all perturbations have
a time-dependence given by $e^{i\o t}$. Also, note that all of the
backgrounds which we have considered have the property
$f_{i,t}=\Phi_{,t}=0$. From the first component of eq.(N3.4), we see that
consistency requires that $h(r)=0$, and that we must also have
$g(t)\sim e^{i\o t}$, which merely results in an additive
constant in $\d f_0$. Similar analyses of the second component of eq.(N3.4)
and of eq.(N3.5) yield no additional constraints. Since eqs. (60) and (75) do
not contain any terms proportional to $\d f_0$ without derivatives, we see
that the  integration constant implicit in eqs. (60) and (75) is actually a
gauge artifact. Thus, after taking account all residual gauge freeedom, one
is left with an uncountably infinite number of distinct solutions to the
linearized equations of motion for both the Witten black hole and the
black hole coupled to a constant, massive dilaton.}

\noindent{We remark briefly on the overall choice of gauge in studying the
small fluctuations of two--dimensional black holes.
One may enquire as to the consequences of choosing conformal gauge in our
analysis, as well as in possible
generalizations of our analysis to other configurations such as the CGHS
black hole. In this case, following the procedure of reference [3], one would
write the metric tensor as $g_{\m\n}=e^{2\rho}\eta_{\m\n}$ with $\rho$ a
scalar
function. In effecting the variation, one must be careful to allow $g_{00}$
and $g_{11}$ to vary {\it independently}. Thus one must take $g_{00}\to
e^{2\rho+2\x}\eta_{00}$ and $g_{11}\to e^{2\rho+2\y}\eta_{11}$. At this point,
it may naively appear to be the case that the residual gauge freedom is fixed
upon choosing $\x=\y$, thereby restoring conformal gauge. In fact, for
general $\x$ and $\y$, and in particular when both are proportional to
$e^{i\o t}$, this cannot be done. It follows from eq.(N3.4) that one has}

$$\x-\y\to\x-\y+g'(t)-h'(r) ~\eqno({\rm N}3.6)$$

\noindent{where the fact that $f_0=f_1$ in conformal gauge has been used. It
is assumed
that $\x-\y\not= 0$ initially. It is clearly not necessary that $\x-\y$ be
equal to the sum of a function of $r$ alone and a function of
$t$ alone. It follows that, in general, functions $g(t)$ and $h(r)$ cannot be
found which are consistent with the restoration of conformal gauge. This is
in particular obvious if, as we require, both $\x$ and $\y$ vary with time as
$e^{i\o t}$.}

\noindent{We finally note that, throughout our analysis, we have made use
of the coordinate system $\pmatrix{x^0&x^1\cr}=\pmatrix{t&r\cr}$ rather
than light cone coordinates. This choice is consistent with our interest in
what occurs outside of the event horizon, as opposed to what occurs throughout
the maximally--extended spacetime.}

\newpage

\sect{{\bf Appendix}}

\noindent{The Christoffel symbols and variations of same which are relevant to
the analysis of the perturbation equations are for convenience recorded
below.}

\noindent{In the general case, one finds:}

$$\G_{00}^0=f_{0,0}~~~~~~\G_{01}^0=f_{0,1}~~~~~~\G_{11}^0=e^{2f_1-2f_0}f_{1,0}$$
$$\G_{00}^1=e^{2f_0-2f_1}f_{0,1}~~~~~~\G_{01}^1=f_{1,0}~~~~~~\G_{11}^1=f_{1,1}
\eqno({\rm A}1)$$
\noindent{and}

$$\d\G_{00}^0=\d f_{0,0}~~~~~~\d\G_{01}^0=\d f_{0,1}~~~~~~\G_{11}^0=
e^{2f_1-2f_0}{\Big [}\d f_{1,0}+2 (\d f_1-\d f_0)f_{1,0}{\Big ]}$$
$$\d\G_{00}^1=e^{2f_0-2f_1}{\Big [}\d f_{0,1}+2(\d f_0-\d f_1)f_{0,1}
{\Big ]}~~~~~~\d\G_{01}^1=\d f_{1,0}~~~~~~\d\G_{11}^1=\d f_{1,1}
\eqno({\rm A}2)$$

\noindent{\bf{The $SL(2,R)/U(1)$ Black Hole}}

\noindent{With the help of eqs. (38) and (39) one may derive the following:}

$$\G_{00}^0=0~~~~~~\G_{01}^0=\sech~\csch~~~~~~\G_{11}^0=0$$
$$\G_{00}^1=\tanh r~\sss~~~~~~\G_{01}^1=0~~~~~~\G_{11}^1=0 ~,\eqno({\rm A}3)$$

$$\d\G_{00}^0=i\o\x~~~~~~\d\G_{01}^0=\x_{,r}~~~~~~\d\G_{11}^0=i\o\chs~\y$$
$$\d\G_{00}^1=-2\tanh r~\sss~\y+2\tanh r~\sss~\x+\tanh^2r\x_{,r}$$
$$\d\G_{01}^1=i\o\y~~~~~~\d\G_{11}^1=\y_{,r} ~.\eqno({\rm A}4)$$

{\bf Massive Dilaton Black Hole}

\noindent{With the help of eq.(66) one may derive the following:}

$$\G_{00}^0=0~~~~~~\G_{01}^0=A^{-1}A'~~~~~~\G_{11}^0=0$$
$$\G_{00}^1={1\over 2}A^2\left(A^2\right)'~~~~~~\G_{01}^1=0~~~~~~\G_{
11}^1={1\over 2}A^2\left(A^{-2}\right)' ~,\eqno({\rm A}5)$$

$$\d\G_{00}^0=i\o\x~~~~~~\d\G_{01}^0=\x_{,r}~~~~~~\d\G_{10}^1=i\o\y$$
$$\d\G_{11}^0=i\o A^{-4}\y~~~~~~\d\G_{00}^1=-A^2{\Big [}\left(A^2\right)'
\y-\left(A^2\right)'\x-A^2\x_{,r}{\Big ]}~~~~~~\d\G_{11}^1=\y_{
,r} ~.\eqno({\rm A}6)$$

\newpage

\noindent{{\bf Table 1}: The Function $\chi(r)$}

\medskip

\vbox{\offinterlineskip
\def\tablerule{\noalign{\hrule}}
\hrule
\halign{&\vrule#&
  \strut\quad\hfil#\quad\cr
height2pt&\omit&&\omit&&\omit&&\omit&&\omit&\cr
&$r$&&$m=1$&&$m=2$&&$m=3$&&$m=4$&\cr\tablerule
height2pt&\omit&&\omit&&\omit&&\omit&&\omit&\cr
&$1.0$&&$-26.9$&&$-1438.5$&&$-18312.6
$&&$-1.35.10^5$&\cr\tablerule
height2pt&\omit&&\omit&&\omit&&\omit&&\omit&\cr
&$2.0$&&$-6.99.10^{3}$&&$-4.42.10^{5}$&&$-5.32.10^6$&&$-3.27.10^{7}$&\cr
height2pt&\omit&&\omit&&\omit&&\omit&&\omit&\cr\tablerule
height2pt&\omit&&\omit&&\omit&&\omit&&\omit&\cr
&$3.0$&&$-8.69.10^{5}$&&$-5.56.10^{7}$&&$-6.43.10^8$&&$-3.70.10^{9}$&\cr
\tablerule
&$4.0$&&$-7.06.10^{7}$&&$-4.52.10^{9}$&&$-5.16.10^{10}$&&$-2.92.10^{11}$&\cr
\tablerule
&$5.0$&&$-4.38.10^{9}$&&$-2.81.10^{11}$&&$-3.20.10^{12}$&&$-1.80.10^{13}$&\cr
\tablerule
&$6.0$&&$-2.27.10^{11}$&&$-1.45.10^{13}$&&$-1.65.10^{14}$&&$-9.30.10^{14}$&\cr
\tablerule
&$7.0$&&$-1.03.10^{13}$&&$-6.60.10^{14}$&&$-7.51.10^{15}$&&$-4.22.10^{16}$&\cr
\tablerule
&$8.0$&&$-4.24.10^{14}$&&$-2.71.10^{16}$&&$-3.09.10^{17}$&&$-1.74.10^{18}$&\cr
\tablerule
&$9.0$&&$-1.61.10^{16}$&&$-1.03.10^{18}$&&$-1.18.10^{19}$&&$-6.61.10^{19}$&\cr
\tablerule
&$10.0$&&$-5.77.10^{17}$&&$-3.69.10^{19}$&&$-4.21.10^{20}$&&$-2.36.10^{21}$&\cr
height2pt&\omit&&\omit&&\omit&&\omit&&\omit&\cr}
\hrule}

\noindent{The above table displays representative values of the function
$\chi(r)$ which arises in the analysis of the small fluctuations of the
two--dimensional black hole coupled to a massive,  linear dilaton, as
discussed above and below eq.(93) in the text.}

\noindent{{\bf Table 2:} The Function $\g(r)$}

\medskip

\vbox{\offinterlineskip
\def\tablerule{\noalign{\hrule}}
\hrule
\halign{&\vrule#&
  \strut\quad\hfil#\quad\cr
height2pt&\omit&&\omit&&\omit&&\omit&&\omit&\cr
&$r$&&$m=1$&&$m=2$&&$m=3$&&$m=4$&\cr\tablerule
height2pt&\omit&&\omit&&\omit&&\omit&&\omit&\cr
&$1.0$&&$1.98$&&$7.26$&&$13.8
$&&$18.29$&\cr\tablerule
height2pt&\omit&&\omit&&\omit&&\omit&&\omit&\cr
&$2.0$&&$14.62$&&$56.41$&&$119.13$&&$192.36$&\cr
height2pt&\omit&&\omit&&\omit&&\omit&&\omit&\cr\tablerule
height2pt&\omit&&\omit&&\omit&&\omit&&\omit&\cr
&$3.0$&&$75.08$&&$297.35$&&$657.87$&&$1141.77$&\cr
\tablerule
&$4.0$&&$327.34$&&$1306.38$&&$2928.15$&&$5177.72$&\cr
\tablerule
&$5.0$&&$1298.41$&&$5191.21$&&$1.17.10^{4}$&&$2.07.10^{4}$&\cr
\tablerule
&$6.0$&&$4841.00$&&$1.94.10^{4}$&&$4.36.10^{4}$&&$7.74.10^{4}$&\cr
\tablerule
&$7.0$&&$1.73.10^{4}$&&$6.91.10^{4}$&&$1.55.10^{5}$&&$2.76.10^{5}$&\cr
\tablerule
&$8.0$&&$5.96.10^{4}$&&$2.38.10^{5}$&&$5.37.10^{5}$&&$9.54.10^{5}$&\cr
\tablerule
&$9.0$&&$2.00.10^{5}$&&$8.02.10^{5}$&&$1.80.10^{6}$&&$3.21.10^{6}$&\cr
\tablerule
&$10.0$&&$6.61.10^{5}$&&$2.64.10^{6}$&&$5.95.10^{6}$&&$1.06.10^{7}$&\cr
height2pt&\omit&&\omit&&\omit&&\omit&&\omit&\cr}
\hrule}

\noindent{The above table displays representative values of the function
$\g(r)$ ({\it cf} eq.(88)) which arises in the analysis of the small
fluctuations of the two--dimensional black hole coupled to a massive, linear
dilaton, as discussed above and below eq.(93) in the text.}
\newpage
\refs

\Item{[1]} P.A.M Dirac, {\it The Principles of Quantum Mechanics}, Clarendon
Press (Oxford), 1958.
\Item{[2]} E.Witten, \pr{D44} (1991) 314.
\Item{[3]} C. Callan, S. Giddings, J. Harvey and A. Strominger, \pr{D45}
(1992) R1005.
\Item{[4]} I. Bars and D. Nemeschansky, \np{348} (1992) 89.
\Item{[5]} M. Rocek, K Schoutens and A. Sevrin, \pl{265} (1991) 303.
\Item{[6]} R. Dijkgraaf, H. Verlinde and E. Verlinde, \np{371} (1992) 269.
\Item{[7]} J. Distler and P. Nelson, \np{366} (1991) 255.
\Item{[8]} S. Chaudhuri and J. Lykken, {\it String Theory, Black Holes and
$SL(2,R)$ Current Algebra}, FERMI-PUB-92/169-T, 6/92.
\Item{[9]} I. Bars, {\it String Propagation on Black Holes}, USC-91/HEP-B3,
5/91; I. Bars and K. Sfetsos, {\it Conformally Exact Metric and Dilaton in
String Theory on Curved Spacetime}, to appear in {\it Phys. Rev.} {\bf D}.
\Item{[10]} P. Ginsparg and F. Quevedo, {\it Strings on Curved Spacetimes:
Black Holes, Torsion, and Duality}, LA-UR-92-640, 2/92.
\Item{[11]} G. Gibbons and M. Perry, {\it The Physics of 2--d Stringy
Spacetimes}, U. Cambridge preprint, 3/92.
\Item{[12]} T. Banks, A. Dabholkar, M. Douglas and
M. O'Loughlin, \pr{D45} (1992) 3607.
\Item{[13]} S.P. DeAlwis, {\it Black Hole Physics from Liouville Theory},
COLO-HEP-284, 6/92.
\Item{[14]} A. Bilal and C. Callan, {\it Liouville Models of Black Hole
Evaporation}, PUPT-1230, 5/92.
\Item{[15]} K .Hamada, {\it Quantum Theory of Dilaton Gravity in $1+1$
Dimensions}, UT-Komaba 92-7, 7/92.
\Item{[16]} S. Giddings and A. Strominger, {\it Quantum Theories of Dilaton
Gravity}, UCSBTH-92-28, 7/92.
\Item{[17]} S. Hawking and J. Stewart {\it Naked and Thunderbolt
Singularities in Black Hole Evaporation}, U. Cambridge preprint, 7/92.
\Item{[18]} J. Russo, L. Susskind and L. Thorlacius, {\it The Endpoint
of Hawking Radiation}, SU-ITP-92-17, 6/92.
\Item{[19]} R. Gregory and J. Harvey, {\it Black Holes with a Massive
Dilaton}, EFI-92-49, 9/92.
\Item{[20]} J. Horne and G. Horowitz, {\it Black Holes Coupled to a Massive
Dilaton}, UCSBTH-92-17, 10/92.
\Item{[21]} M. McGuigan, C. Nappi and S. Yost, {\it Charged Black Holes in
Two--Dimensional String Theory}, IASSNS-HEP-91/57, 10/91.
UCSBTH-92-17, 9/92.
\Item{[22]} G. Mandal, A. Sengupta and S. Wadia, {\it Mod. Phys.
Lett} {\bf A6} (1991) 1685.
\Item{[23]} T. Buscher, \pl{201} (1988) 466.
\Item{[24]} T. Damour, G. Gibbons and C. Gundlach, \prl{64}, (1990) 123.
\Item{[25]} S. Chandrasekhar, {\it Proc. Roy. Soc.} {\bf A365} (1979) 453.
\Item{[26]} G. Curci and G. Pafutti, \np{312} (1989) 227.
\Item{[27]} G. Gilbert, {\it On the Perturbations of String Theoretic Black
Holes}, UMDEPP 92-035 (revised version), 8/92.
\Item{[28]} S.P. de Alwis and J. Lykken, \pl{269} (1991) 264.
\Item{[29]} G. Diamandis, B. Georgalas, X. Maintas, N. Mavromatos, {\it
Time-dependent Perturbations in Two-Dimensional String Black Holes},
CERN-TH 6671/92, 9/92.
\end{document}